# Dynamically Reconfigurable Optical Skyrmions Enabled by a Silicon Microring Optical Phased Array for Robust Free-Space Communication

Zili Cai, Tian Zhang*, Qi Chen, Jubo Hao, Jian Dai, and Kun Xu
*State Key Laboratory of Information Photonics and Optical Communications, Beijing University of Posts and Telecommunications, Beijing 100876, China*

**ABSTRACT:** Optical skyrmions can encode information in the skyrmion number for free-space communication through atmospheric turbulence. However, programming both the skyrmion number and type on a compact chip remains challenging. Here, we propose and numerically demonstrate a skyrmion generator based on a silicon optical phased array with optimized inner- and outer-grating microring emitters. The two emitter types provide orthogonal circularly polarized components and generate optical skyrmions through coherent superposition. The on-chip phase shifters configure the azimuthal phase distribution, enabling the 4×4 array to generate states with target skyrmion numbers of −2, −1, 1, and 2. Adjustment of the relative phase bias continuously switches states with target skyrmion numbers of ±1 between the Néel and Bloch types. Increasing the array size to 12×12 extends the target range from −8 to 8. Laser wavelength tuning and calibration of the microring input phases and amplitudes recover target states within a range of simulated fabrication errors, which shows the robustness of the generator. Furthermore, we evaluate four-symbol communication using skyrmion-number decoding over a simulated 100 m turbulent link and observe no symbol errors in $10^4$ trials at Rytov variances up to 0.12. These results demonstrate the potential of this generator as a compact, programmable transmitter for turbulence-tolerant free-space optical communication.

## 1. INTRODUCTION

Driven by the rapid development of data-intensive applications such as artificial intelligence and cloud computing, next-generation optical communication systems demand higher capacity and stronger robustness against channel disturbances.[1,2] Conventional multiplexing techniques have been the primary route to capacity scaling, but they are now approaching their physical limits.[3–5] Structured light carrying orbital angular momentum (OAM) has emerged as a promising route to further increase the communication capacity, but its performance is still limited by channel disturbances, mainly the phase distortion induced by atmospheric turbulence.[6–9] Optical skyrmions, an emerging type of vectorial information carrier, have the potential to meet both requirements, namely high capacity and robustness, at once.[10–13] Across the beam cross section of an optical skyrmion, the polarization state varies continuously and wraps the Poincaré sphere non-trivially, and its topological texture, characterized by the skyrmion number, remains intact under smooth perturbations.[10,14–16] Such a texture can be generated by coherently superposing OAM modes of different orders with orthogonal circular polarizations,[14–16] so that skyrmion-based channels can inherit the capacity advantage of OAM multiplexing while adding topological robustness against channel disturbances.[10,11,17–19]

Although optical skyrmions have broad application prospects, generating them efficiently in practical systems remains difficult. The most common route relies on conventional free-space optical systems, typically built with spatial light modulators[20–22] or q-plates.[23,24] These systems provide flexible and programmable control over the generated optical skyrmions,[22] but their large footprints and strict alignment requirements severely constrain their practical deployment. More compact alternatives have therefore been pursued. Optical metasurfaces[25,26] are used to shrink the functional element to the subwavelength scale and to generate optical skyrmions with precisely designed textures. In addition, on-chip generators integrate the functional elements on a photonic chip and further reduce the footprint of the system. For example, in silicon microring resonators (MRRs) with double angular gratings, two orthogonally polarized beams are diffracted and superposed into a skyrmion beam above the chip.[27] However, in these two kinds of compact generators, the skyrmion texture is largely fixed by the fabricated structure, so the dynamic and flexible control of the generated optical skyrmions is still limited.

Recently, switching between different skyrmion types has been explored. A plasmonic platform uses binary PIN-diode coding to alternate between a Néel-type skyrmion and a meron,[28] but it offers only two discrete states. More recently, continuous programming of the skyrmion texture between Néel and Bloch types has been demonstrated on a silicon photonic chip,[29] but only skyrmions with $|N_{sk}|=1$ are demonstrated. Tuning of the skyrmion number has also been pursued. Nonlinear Čerenkov radiation in thin-film lithium-niobate MRRs has been exploited to generate reconfigurable optical skyrmions,[30] but in each configuration the skyrmion number is locked to one preset discrete order (±2 or ±4), which is determined by the nonlinear process and the pump-mode combination. Therefore, a single-chip platform that programs the skyrmion number over a wide range of orders and also switches the skyrmion type at telecommunication wavelengths is still lacking.

In this work, we propose SIGMA (skyrmion generator based on a microring optical phased array), a dynamically reconfigurable optical skyrmion generator based on a silicon optical phased array (OPA) for robust free-space communication. SIGMA integrates inner-grating and outer-grating MRR emitters (IG-MRRs and OG-MRRs), which generate two orthogonal circular polarization states that are independently controlled and coherently superposed. First, the angular gratings of the IG-MRRs and OG-MRRs are optimized by the particle swarm optimization (PSO) algorithm, so that the emitted beams are left- and right-circularly polarized (LCP and RCP) with polarization fractions of 88.08% and 93.68%, respectively. Second, by controlling the phase bias $\Delta\delta$ between the input phases of the IG-MRRs and OG-MRRs, the skyrmions with target $N_{sk}=\pm1$ can be tuned continuously between the Néel and Bloch types without changing $N_{sk}$. By controlling the difference of the OAM order $\Delta l$ between the LCP and RCP components, the skyrmion number can be programmed over a wide range. We have numerically demonstrated that a compact 4×4 OPA can tune the skyrmion number $N_{sk}$ from −1.925 to 1.902, and an enlarged 12×12 OPA can cover target skyrmion numbers from −8 to 8. We further construct a four-symbol skyrmion-encoded free-space optical communication channel over 100 m of atmospheric turbulence. The skyrmion-encoded channel shows no symbol errors in $10^4$ trials up to a Rytov variance of 0.12, whereas the symbol error rate of a Laguerre–Gaussian OAM (LG-OAM) channel under the same turbulence already reaches 0.60. This work turns the optical skyrmion from a statically fabricated pattern into a dynamically programmable information carrier and offers a compact single-chip route toward high-capacity, turbulence-tolerant free-space optical communication.

## 2. DEVICE DESIGN AND THEORETICAL MODEL

### 2.1. System Configuration

The proposed SIGMA is built on a silicon chiral OPA, as shown in Figure 1. The system consists of two parts, namely a modulation network formed by a power distribution network and a phase shifter array, and a two-dimensional $N \times N$ emitter array formed by chiral MRRs. A continuous-wave input is routed through these two parts in sequence and is finally converted into an OAM beam with a vectorial polarization texture in free space. The system is designed on a silicon photonic platform and operates at the telecommunication wavelength of $\lambda_0$=1550 nm. The power distribution network is a cascaded tree structure formed by tunable Mach–Zehnder interferometer (MZI) splitters.[31,32] By tuning the splitting ratio of each MZI, the network distributes the input light into $N^2$ waveguide channels with programmable weights of the electric-field amplitude. The phase shifter array then independently imposes programmable phase delays on the guided transverse electric (TE) modes. Each unit of the two-dimensional $N \times N$ emitter array is a silicon MRR. The waveguides of the MRRs have a width of $W_{wg}$=450 nm and a height of $H_{wg}$=340 nm. Each MRR is coupled to a bus waveguide with a width of $W_{bus}$=345 nm through a pulley coupler, in which the bus waveguide wraps around the MRR over an angle of 90°. The coupling gap is 200 nm for the IG-MRRs and 150 nm for the OG-MRRs.

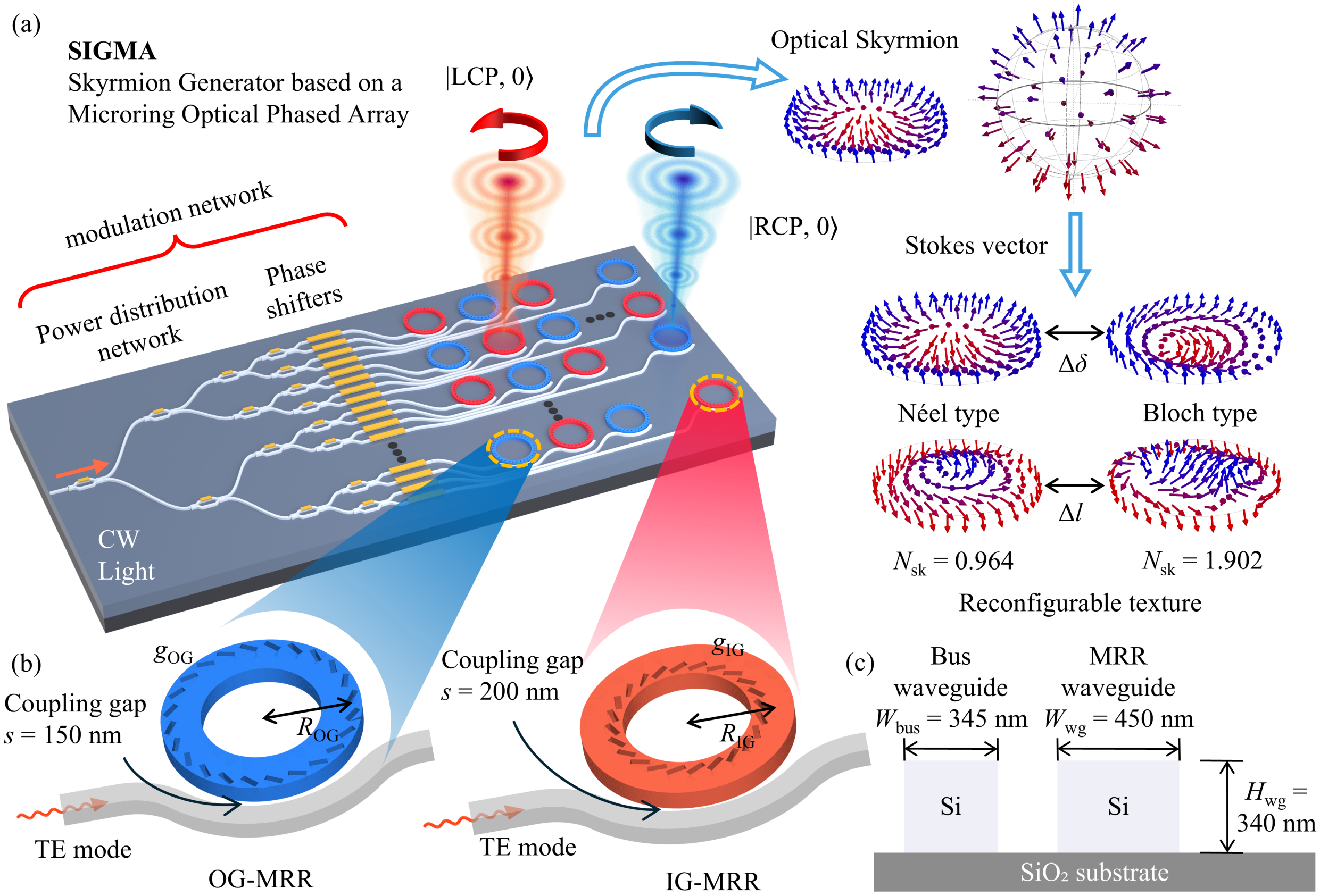

Figure 1. Schematic of the on-chip dynamically reconfigurable optical skyrmion generator. (a) The integrated platform uses a modulation network to address a two-dimensional checkerboard array of chiral MRRs on a silicon OPA. The IG-MRRs and OG-MRRs emit the LCP and RCP components with zero OAM order, which synthesize the far-field polarization texture mapped onto the Poincaré sphere. The phase bias $\Delta\delta$ switches the skyrmion type, and $\Delta l$ changes the skyrmion number. (b) OG-MRR and IG-MRR with outer and inner angular gratings and pulley couplers. (c) Cross section of the bus and MRR waveguides.

The emitter array consists of two types of chiral MRRs, namely the IG-MRRs and the OG-MRRs,

and each emitter supports a quasi-TE whispering gallery mode (WGM) near $\lambda_0$. The two types of MRRs are interleaved in a checkerboard pattern and form two subarrays that are independently controlled by the modulation network. For each type of emitter, the tooth number of the angular grating is designed separately, so that the dominant circularly polarized component of the emitted beam carries zero-order OAM. The geometry of the angular grating is also optimized separately for each type, so that the emitted beam has a high circular polarization purity, as described in Section 2.2. A higher circular polarization purity leaves less residual light in the non-target circular polarization and therefore allows a more precise tuning of the synthesized skyrmion.[27]

### 2.2. Optimization of the Chiral MRR Emitters

The geometry of the angular grating of each MRR emitter is optimized by a particle swarm optimization (PSO) algorithm, as shown in Figure 2a.

1. Initialization of the optimization parameters. In the PSO framework, each particle is described by five geometric parameters, namely the etch depth ($d_g$), grating length ($l_g$), grating width ($w_g$), grating offset ($o_g$), and the grating tilt angle ($\theta_g$). During the PSO initialization, the particles are randomly distributed within the predefined search ranges: 30–100 nm for $d_g$, 60–350 nm for $l_g$, 40–130 nm for $w_g$, 0–180 nm for $o_g$, and −75° to 75° for $\theta_g$. The initial velocity of each particle is set to 0, the swarm size is 20, and the maximum number of generations is 50. For each type of MRR, the PSO runs three times with different random initializations.
2. Calculation of the tooth numbers of the angular gratings. When a WGM with azimuthal mode number $m$ is scattered by an angular grating with $g$ teeth, the radiated beam in the far field carries an azimuthal order of $l_0$. The electric field $\boldsymbol{E}(\theta, \varphi)$ of the radiated OAM beam can be decomposed into LCP and RCP components with different azimuthal orders[33]:
$$\mathbf{E}(\theta,\varphi) = c_{\mathrm{L}}(\theta)e^{i(l_0-1)\varphi}\hat{\mathbf{e}}_{\mathrm{LCP}} + c_{\mathrm{R}}(\theta)e^{i(l_0+1)\varphi}\hat{\mathbf{e}}_{\mathrm{RCP}} \quad (1)$$
where $\theta$ and $\varphi$ are the polar and azimuthal angles of the emission direction. $c_{\mathrm{L}}(\theta)$ and $c_{\mathrm{R}}(\theta)$ are the complex amplitudes of the LCP and RCP components, respectively, $e^{i(l0-1)\varphi}$ and $e^{i(l0+1)\varphi}$ denote the azimuthal phase factors of the two components, and $\hat{e}_{LCP}$ and $\hat{e}_{RCP}$ are the unit polarization vectors of the LCP and RCP light. Eq. (1) shows that the LCP component carries an OAM order of $l_0-1$ while the RCP component carries an OAM order of $l_0+1$. Notably, in the far field, the power of the circularly polarized component with zero OAM order is concentrated in the central spot, which is also the region where the optical skyrmion is generated, as described in Section 2.3. Therefore, in this work, the target circularly polarized component emitted by each MRR, namely the LCP component from the IG-MRR and the RCP component from the OG-MRR, is designed to carry zero OAM order. Specifically, for the IG-MRR, the azimuthal order $l_{0,\mathrm{IG}}$ is set to 1 so that the target LCP component carries zero OAM order, and for the OG-MRR, the azimuthal order $l_{0,\mathrm{OG}}$ is set to −1. To realize the required azimuthal order, the tooth number $g$ of the angular grating must satisfy[34]:
$$g = m - l_0 \quad (2)$$
Considering the footprint of the system, the initial radii of the IG-MRR and the OG-MRR, $R_{\mathrm{IG}}$ and $R_{\mathrm{OG}}$, are both set to 3.2 μm, and the WGM with the azimuthal mode number $m$=33 is used. The resonant wavelength of this WGM follows the resonance condition $2\pi R\cdot n_{\mathrm{eff}}=m\lambda_0$,[35] where $n_{\mathrm{eff}}$ is the effective index of the WGM in the MRR without the angular grating. Since adding the angular grating to the MRR slightly changes $n_{\mathrm{eff}}$, the radius $R$ is fine-tuned in Step 8 to align the resonance of this WGM with $\lambda_0$. According to Eq. (2), the tooth number of the angular grating is $g_{IG}=m-l_{0,IG}=32$ for the IG-MRR and $g_{OG}=m-l_{0,OG}=34$ for the OG-MRR.

3. Setting of the fitness score. The fitness score is defined by the degree of circular polarization (DOCP), the upward emission efficiency $\eta_{up}$, the back reflectance $R_{back}$, and the direction ratio $D$. The fitness score is then calculated as:

$$F = I_{pol} \left|DOCP\right|^{\gamma} \exp(-\kappa R_{back}) P_{\eta}(\eta_{up}) D \tag{3}$$

where $I_{pol}$ is a selector that equals 1 when the sign of the DOCP matches the target polarization (DOCP>0 for the LCP target and DOCP<0 for the RCP target) and equals 0 otherwise. $\eta_{up}$ is the power flowing upward through a plane above the angular grating, normalized by the power of the injected TE mode. The DOCP is defined as the normalized power difference between the LCP and RCP components over the far field[36]:

$$DOCP = \frac{P_{LCP} - P_{RCP}}{P_{LCP} + P_{RCP}} \tag{4}$$

where $P_{LCP}$ and $P_{RCP}$ are the powers of the LCP and RCP components radiated into the far field above the grating, and their calculation is described in Step 5. The exponent $\gamma$ in Eq. (3) determines the sensitivity of the fitness score to the DOCP, and we set $\gamma$=8. The term $\exp(-\kappa R_{back})$ is a penalty factor for the back-reflection, and $\kappa$ determines the strength of this penalty, which is set to 50. The direction ratio $D=\eta_{up}/(\eta_{up}+\eta_{down})$ is the fraction of the vertically radiated power that goes upward, where $\eta_{down}$ is the power flowing downward into the substrate below the angular grating, normalized by the power of the injected TE mode. The term $D$ favors the angular gratings that radiate more power upward. The efficiency term $P_{\eta}$ is a flat-top window function. It equals 1 when $\eta_{up}$ lies within the target window and decreases with a Gaussian profile outside it. The upper bound of $\eta_{up}$ is derived from the loaded quality factor $Q_L$ and we set $Q_L$=1200 to balance the upward emission efficiency and the polarization purity of the emitted beam.[33] For an MRR with radius $R$ ($R=R_{IG}$ for the IG-MRR and $R=R_{OG}$ for the OG-MRR), group index $n_g$ of the TE mode and operating wavelength $\lambda_0$, the round-trip loss $\delta_{rt}$ is[35]:

$$\delta_{rt} = \frac{4\pi^2 R n_g}{\lambda_0 Q_L} \tag{5}$$

Since the three-period model contains three of the $g$ grating periods and the upward emission is only one part of the round-trip loss, we conservatively limit $\eta_{up}$ to less than half of the share of $\delta_{rt}$ in the three periods:

$$\eta_{up} \leq \frac{3}{2g} \delta_{rt} \tag{6}$$

The target window of $P_{\eta}$ is set to $0.1\% \leq \eta_{up} \leq 0.5\%$, which lies within the bound given by Eq. (6) for both MRRs.

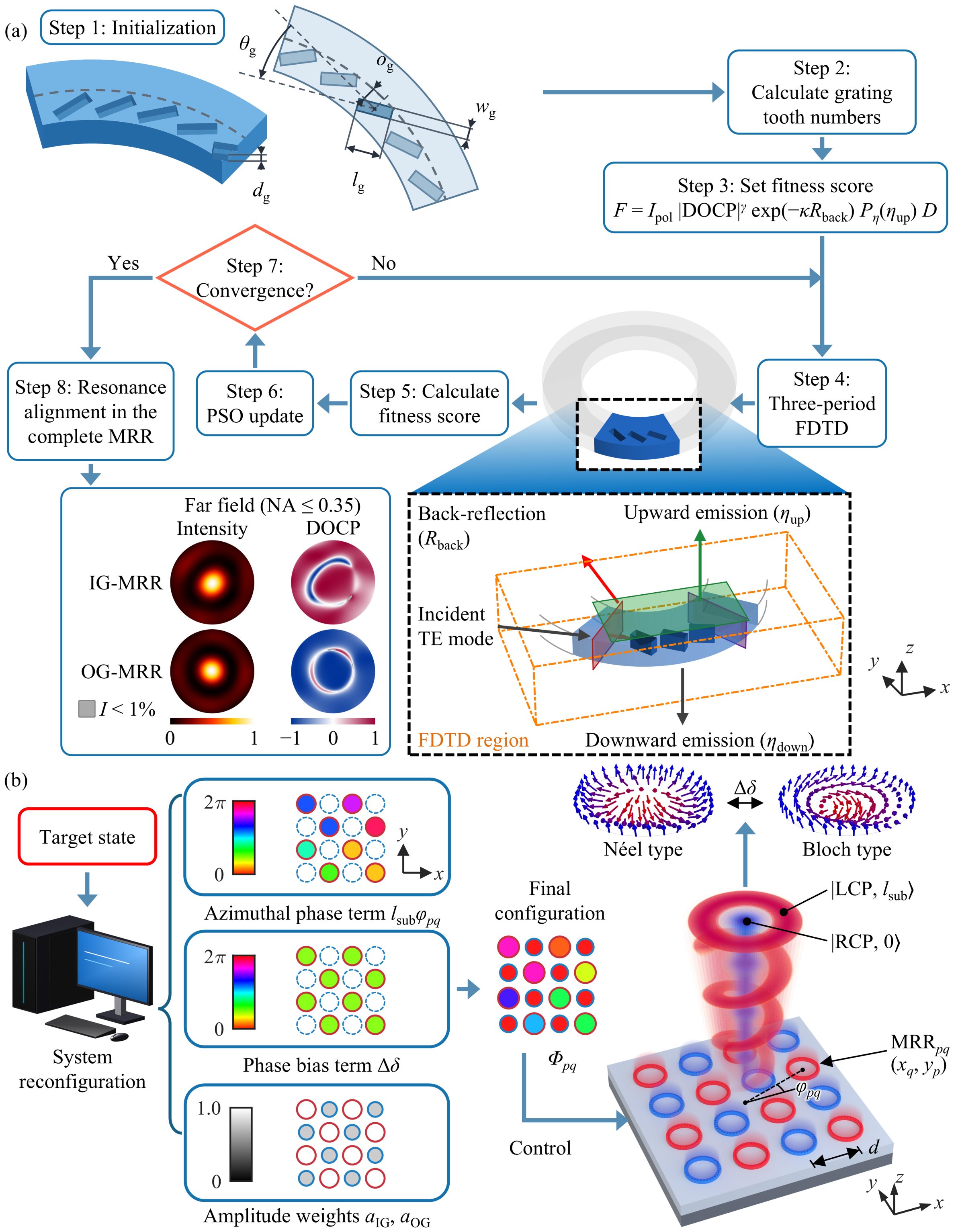

Figure 2. Optimization of the MRR emitters and the OPA synthesis workflow. (a) Optimization of the chiral MRR emitters. The grating geometry is optimized by particle swarm optimization with a three-period FDTD model, and the fitness score includes the circular polarization purity, the upward emission efficiency, the back-reflection, and the fraction of the vertically radiated power

that goes upward. After the resonance alignment, the optimized parameters are validated by far-field simulations of the complete MRR with the pulley coupler at resonance. (b) OPA synthesis of optical skyrmions. The optimized IG-MRR and OG-MRR emitters form two orthogonal circular polarization bases. The programmable phase maps and amplitude weights of the OPA are applied to the checkerboard emitter array to control the OAM orders and the relative phase between the LCP and RCP components, thereby generating reconfigurable skyrmion states. The illustrated configuration applies the azimuthal phase with $l_{\mathrm{sub}}=-1$ and the phase bias to the IG-MRRs. Red and blue outlines mark the IG-MRRs and the OG-MRRs, and dashed circles mark emitters that do not receive the term. The coaxial RCP and LCP components form the core and the ring of the beam.

4. Three-period FDTD simulation. Each particle is evaluated by the three-dimensional finite-difference time-domain (3D-FDTD) method on a local model that contains only three angular grating periods of the curved waveguide section,[37] as shown in Figure 2a. A fundamental TE mode is injected into the curved waveguide section, and the etched angular grating scatters part of the guided field into free space. Since the angular grating is periodic along the ring, the guided mode in the grating region is a Bloch mode, whose field profile is the same in every period except for a phase factor.[38] The polarization of the light scattered by each period is determined by the radial and azimuthal electric fields of this mode,[33,39] so a local model containing only three grating periods can approximately reproduce the emission polarization of the complete MRR at a substantially reduced simulation cost.[37]
5. Fitness score calculation. After the 3D-FDTD simulation of the three-period model, we project the field in the *x*-*y* plane to the far field and integrate the upward component of the far-field Poynting vector over the upper hemisphere in spherical coordinates:
$$P_\sigma = \frac{r^2}{2\eta_0}\int_0^{2\pi}\int_0^{\frac{\pi}{2}}\left|E_\sigma(r,\theta,\varphi)\right|^2 \cos\theta\sin\theta d\theta d\varphi \tag{7}$$
where $\sigma\in\{\mathrm{LCP},\mathrm{RCP}\}$ denotes the circularly polarized component, $\eta_0$ is the free-space impedance, and $E_{\mathrm{LCP}}=(E_x-iE_y)/\sqrt{2}$ and $E_{\mathrm{RCP}}=(E_x+iE_y)/\sqrt{2}$ are the far-field electric fields of the LCP and RCP components. Then, by substituting $P_\sigma$ into Eqs. (4) and (3), the fitness score $F$ is obtained.
6. PSO update. After all the particles in the current generation are evaluated, we update the personal best and the global best solutions based on the obtained fitness scores of all particles. Then, the velocities and positions of each particle are adjusted according to the PSO algorithm,[40] and the updated geometric parameters $\{d_g, l_g, w_g, o_g, \theta_g\}$ are evaluated in the next generation.
7. Convergence check. If the maximum number of generations has not been reached, Steps 4–7 are repeated. For each MRR type, the geometry with the highest fitness score among the three independent PSO runs is selected.
8. Resonance alignment in the complete MRR. By simulating the complete MRR with the optimized angular grating using the 3D-FDTD method, we obtain the accurate $n_{\mathrm{eff}}$,[41] which is slightly different from the initial value used in Step 2. By fine-tuning the radius $R$ of the MRR with the optimized angular grating, the WGM with the azimuthal mode number $m$ resonates again at the operating wavelength $\lambda_0$.[42] Notably, the optimized geometric parameters of the angular grating are not modified in this step, so that the optimized polarization purity is preserved.[39]

### 2.3. Synthesis and Programming of Far-Field Optical Skyrmions

In Section 2.2, the IG-MRR and the OG-MRR are designed to emit an LCP beam and an RCP beam, respectively, and the target circularly polarized component of each beam carries zero OAM order. In this section, we synthesize these two beams into an optical skyrmion with SIGMA.[43] The input phases of the emitter array add an OAM order of $l_{\mathrm{LCP}}$ to the LCP beam emitted by the IG-MRRs and an OAM order of $l_{\mathrm{RCP}}$ to the RCP beam emitted by the OG-MRRs. Here, the residual circularly polarized component of the beam emitted by each MRR is neglected, so that in this ideal model, the IG-MRR emits only LCP light and the OG-MRR emits only RCP light. Under the paraxial approximation, the far field of the skyrmion beam synthesized from these two circularly polarized components can be written as:

$$\begin{aligned}\mathbf{E}_{\mathrm{SIGMA}}(\theta,\varphi) &= a_{\mathrm{IG}}\, b_{\mathrm{IG}}(\theta) e^{i(l_{\mathrm{LCP}}\varphi+\delta)}\, \hat{\mathbf{e}}_{\mathrm{LCP}} \\ &+ a_{\mathrm{OG}}\, b_{\mathrm{OG}}(\theta) e^{i l_{\mathrm{RCP}}\varphi}\, \hat{\mathbf{e}}_{\mathrm{RCP}}\end{aligned} \tag{8}$$

where $a_{\mathrm{IG}}$ and $a_{\mathrm{OG}}$ are the input amplitudes of the IG-MRRs and the OG-MRRs, respectively. $b_{\mathrm{IG}}(\theta)$ and $b_{\mathrm{OG}}(\theta)$ describe the radial amplitudes of the two subarray fields by multiplying $c_{\mathrm{L}}(\theta)$ of the IG-MRR and $c_{\mathrm{R}}(\theta)$ of the OG-MRR in Eq. (1) by their respective array factors. $\delta$ is the phase difference of the LCP component relative to the RCP component. The OAM-order difference $\Delta l=l_{\mathrm{LCP}}-l_{\mathrm{RCP}}$ determines the skyrmion number $N_{\mathrm{sk}}$,[14,43] and the phase difference $\delta$ determines the skyrmion type.[10]

To program the skyrmion number $N_{\mathrm{sk}}$ and the skyrmion type to our target, we need to decide the phase configuration of the modulation network in our SIGMA. In our proposed SIGMA, the IG-MRRs and the OG-MRRs are arrayed into an $N\times N$ checkerboard and lie in the $x$-$y$ plane with the origin at its center. As shown in Figure 2b, the MRR in the $p$-th row and the $q$-th column is denoted as MRR$_{pq}$, and MRR$_{pq}$ is located at $(x_q, y_p)$, with $x_q=(q-(N+1)/2)d$ and $y_p=((N+1)/2-p)d$, and its azimuthal position with respect to the array center is $\varphi_{pq}=\mathrm{atan2}(y_p, x_q)$. MRR$_{pq}$ is an IG-MRR when $p+q$ is even and an OG-MRR when $p+q$ is odd. By tuning the phase shifters in the power distribution network, we program the input amplitudes $a_{\mathrm{IG}}$ and $a_{\mathrm{OG}}$. Since the optimized IG-MRR and OG-MRR radiate different powers into the main lobe, the amplitude ratio $a_{\mathrm{IG}}/a_{\mathrm{OG}}$ is set for each target state to control the relative strength of the LCP and RCP components in Eq. (8).

The OAM orders $l_{\mathrm{LCP}}$ and $l_{\mathrm{RCP}}$ and the phase difference $\delta$ in Eq. (8) are controlled by the input phase delay $\Phi_{pq}$ of MRR$_{pq}$, and $\Phi_{pq}$ is programmed by the phase shifter array in SIGMA as below:

$$\Phi_{pq} = \begin{cases} l_{\mathrm{LCP}}\varphi_{pq} + \Delta\delta, & \text{IG-MRR} \\ l_{\mathrm{RCP}}\varphi_{pq}, & \text{OG-MRR} \end{cases} \tag{9}$$

where $l_{\mathrm{LCP}}\varphi_{pq}$ and $l_{\mathrm{RCP}}\varphi_{pq}$ are the azimuthal phases applied to the IG-MRRs and the OG-MRRs, respectively. These azimuthal phase distributions introduce constant phase shifts of $-|l_{\mathrm{LCP}}|\pi/2$ and $-|l_{\mathrm{RCP}}|\pi/2$ into the far-field LCP and RCP components, respectively.[42] $\Delta\delta$ is a constant phase bias applied to all IG-MRRs. Thus the relative phase $\delta$ can be calculated as:

$$\delta = \Delta\delta + \delta_{\mathrm{MRR}} - \frac{(|l_{\mathrm{LCP}}| - |l_{\mathrm{RCP}}|)\pi}{2} \tag{10}$$

where $\delta_{\mathrm{MRR}}$ is the phase difference between the LCP component emitted by the IG-MRR and the RCP component emitted by the OG-MRR near the beam center, which is obtained from their far fields in Section 3.1.

The skyrmion number $N_{\mathrm{sk}}$ is defined as the number of times that the normalized Stokes vector wraps the Poincaré sphere over the beam cross section.[10] SIGMA programs the azimuthal phase on only one of the two subarrays, i.e., $(l_{\mathrm{LCP}}, l_{\mathrm{RCP}})=(l_{\mathrm{sub}}, 0)$ or $(0, l_{\mathrm{sub}})$, where $l_{\mathrm{sub}}$ is the nonzero azimuthal phase order. The magnitude of $N_{sk}$ is equal to $|\Delta l|$[14,43] and $N_{sk}$ has the same sign as $\Delta l$ when the RCP component is at the beam center ($l_{RCP}$=0) and the opposite sign when the LCP component is at the beam center ($l_{LCP}$=0). In both cases, $\Delta l = l_{sub}$ for ($l_{sub}$, 0) and $\Delta l = -l_{sub}$ for (0, $l_{sub}$), and $N_{sk}$ can be calculated as:

$$N_{sk} = l_{sub} \tag{11}$$

We determine the skyrmion type from the orientation of the in-plane Stokes vector, which depends on the relative phase $\delta$ in Eq. (8). The Stokes parameters are $S_0=|E_x|^2+|E_y|^2$, $S_1=|E_x|^2-|E_y|^2$, $S_2=2\mathrm{Re}(E_x^*E_y)$, and $S_3=2\mathrm{Im}(E_x^*E_y)$, where $E_x$ and $E_y$ are the components of $\mathbf{E}_{SIGMA}(\theta, \varphi)$ along the $x$ and $y$ axes. Substituting Eq. (8) into these definitions gives $S_1+iS_2=2a_{IG}a_{OG}b_{IG}b_{OG}\exp[-i(\Delta l\varphi+\delta)]$. The in-plane Stokes vector ($S_1$, $S_2$) therefore points at an angle of $-(\Delta l\varphi+\delta)$ from the $x$ axis. Since the local radial direction is at an angle of $\varphi$, the angle between ($S_1$, $S_2$) and the local radial direction is $\chi=-(\Delta l+1)\varphi-\delta$. The beam is a Néel-type optical skyrmion when $\chi$ is 0 or π, a Bloch-type optical skyrmion when $\chi$ is ±π/2, and a hybrid-type optical skyrmion for other values.[10] When $\Delta l=-1$, $\chi$ is equal to $-\delta$ at every azimuthal position. This case includes the two states ($l_{LCP}$, $l_{RCP}$)=(−1, 0) and (0, 1), which have $N_{sk}=-1$ and +1, respectively. For these two configurations, Eq. (10) gives $\delta=\Delta\delta+\delta_{MRR}-\pi/2$ and $\delta=\Delta\delta+\delta_{MRR}+\pi/2$, respectively. The Néel type ($\delta$=0 or π) is obtained at $\Delta\delta=\pi/2-\delta_{MRR}+j\pi$, and the Bloch type ($\delta=\pm\pi/2$) is obtained at $\Delta\delta=-\delta_{MRR}+j\pi$, where $j$ is an integer. For other values of $\Delta l$, $\chi$ varies with $\varphi$, and the texture is neither of the Néel type nor of the Bloch type. For example, the state with ($l_{LCP}$, $l_{RCP}$)=(1, 0) has $\Delta l$=1 and, according to Eq. (11), $N_{sk}$=+1, but its in-plane Stokes vector winds in the opposite direction and forms a saddle-like anti-skyrmion texture.[10]

In summary, the azimuthal phase order $l_{sub}$ programs the skyrmion number $N_{sk}$, the amplitude ratio $a_{IG}/a_{OG}$ sets the relative strength of the LCP and RCP components, and the phase bias $\Delta\delta$ switches the skyrmion type for the two configurations with $\Delta l=-1$. All three are programmed by the modulation network without changing the MRRs, as illustrated in Figure 2b.

## 3. RESULTS

### 3.1. Optimized Chiral MRR Emitters

Table 1 lists the optimized geometric parameters of the IG-MRR and the OG-MRR which are obtained from Section 2.2. As shown in Figure 3a, the fitness scores converge within about 20 generations, and the geometric parameters with the best fitness score are adopted for each MRR type. Then, as described in Step 8 of Section 2.2, the radii of the MRRs are fine-tuned from the initial value of 3.2 μm to 3.173 μm for the IG-MRR and to 3.165 μm for the OG-MRR, so that the two MRRs resonate at 1550.11 nm and 1549.77 nm, respectively. As shown in Figure 3b, the upward emission efficiencies of the IG-MRR and the OG-MRR are 13.4% and 25.1%, and their loaded quality factors are 840 and 708, respectively.

In the far field, the IG-MRR and the OG-MRR emit two beams with orthogonal circular polarizations and similar intensity profiles in the ($k_x/k_0$, $k_y/k_0$) plane, as shown in Figure 3c. The phase distributions in the insets of Figure 3c further indicate that the dominant circularly polarized component of each emitter carries zero OAM order, as designed in Step 2 of Section 2.2. In Figure 3d, the two emitters exhibit opposite DOCP signs over most of the radiation region, and the target circular polarization fractions of the IG-MRR (LCP) and the OG-MRR (RCP) reach 88.08% and 93.68%, respectively. Here, the target circular polarization fraction is the ratio of the power of the target circularly polarized component to the total power, both integrated over the main lobe ($\theta\leq 20.5°$, i.e., NA≤0.35). Therefore, the IG-MRR and the OG-MRR provide two orthogonal circular polarization bases with high purity for the synthesis of optical skyrmions.

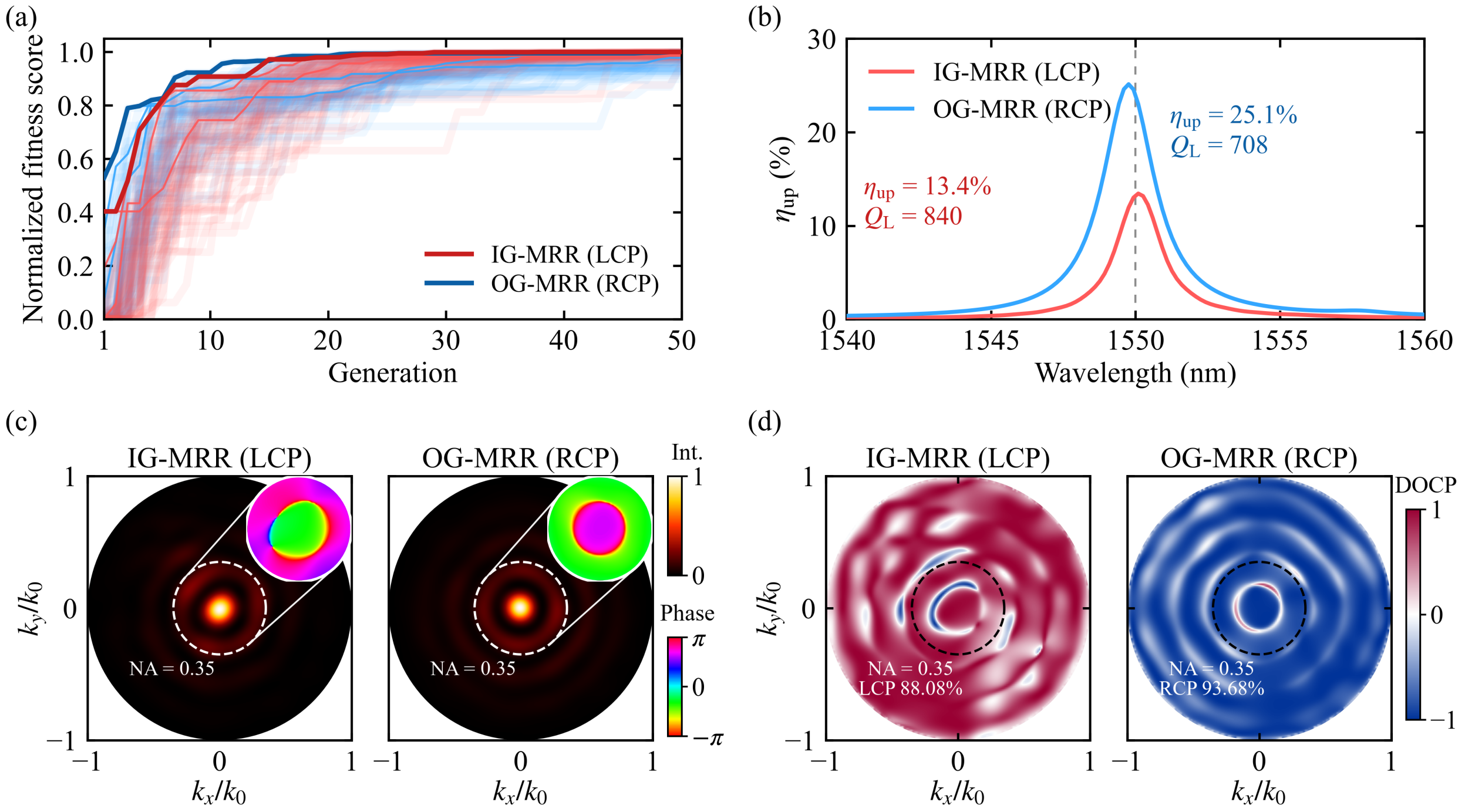

Figure 3. Optimization and complete-MRR FDTD validation of the IG-MRR and OG-MRR emitters. (a) Fitness score of the three-period PSO versus generation, normalized by the global best of each MRR type, for three independent runs of each MRR type. Light lines, thin lines, and bold lines show the best score of each particle, the best of each run, and the global best, respectively. (b) Upward emission efficiency spectra of the complete MRRs. (c) Far-field intensity profiles in the ($k_x/k_0$, $k_y/k_0$) plane. The insets show the phase of the target circularly polarized component within NA≤0.35 (dashed circle). (d) Far-field DOCP ($S_3/S_0$) distributions. Within NA≤0.35 (dashed circles), the LCP fraction of the IG-MRR is 88.08%, and the RCP fraction of the OG-MRR is 93.68%.

Table 1. Geometric Parameters of the IG-MRR and OG-MRR Emitters

| Parameters | IG-MRR | OG-MRR |
|---|---|---|
| Grating length $l_g$ (nm) | 350.0 | 344.2 |
| Grating width $w_g$ (nm) | 129.1 | 61.6 |
| Etch depth $d_g$ (nm) | 77.3 | 96.8 |
| Grating offset $o_g$ (nm) | 111.2 | 124.6 |
| Grating tilt angle $\theta_g$ (deg) | +23.8 | −28.4 |
| Grating tooth number $g$ | 32 | 34 |
| Ring radius $R$ (nm) | 3172.71 | 3164.63 |
| Coupling gap $s$ (nm) | 200 | 150 |

### 3.2. Phase-Controlled Switching of the Skyrmion Type

Here, we use a 4×4 SIGMA with a pitch of 10 μm to verify the switching between Néel-type and Bloch-type skyrmions. The far field of this SIGMA is calculated by superposing the far fields of the IG-MRR and the OG-MRR obtained in Section 3.1, with the input amplitudes $a_{IG}$ and $a_{OG}$ and the input phase delay $\Phi_{pq}$ in Eq. (9). According to Section 2.3, the phase bias $\Delta\delta$ switches the skyrmion type in the two states with $\Delta l=-1$, namely ($l_{\mathrm{LCP}}$, $l_{\mathrm{RCP}}$)=(−1, 0) with target $N_{\mathrm{sk}}=-1$ and (0, 1) with target $N_{\mathrm{sk}}=+1$. For these two states, the amplitude ratio $a_{\mathrm{IG}}/a_{\mathrm{OG}}$ is set to 1.52 and 1.02, respectively, and the corresponding upward emission efficiencies of the whole 4×4 array are about 17% and 19%. Then, only the phase bias $\Delta\delta$ is swept from 0 to 2π with 40 sampling points for each state.

At each phase bias $\Delta\delta$, the skyrmion number and the skyrmion type of the synthesized far field are evaluated in the observation region, which contains the emission directions within NA≤0.04. This region approximately covers the main lobe of the 4×4 array, and the same observation region is used for all the arrays in this work. Firstly, $N_{\mathrm{sk}}$ is calculated by integrating the skyrmion number density of the normalized Stokes vector over the observation region:

$$N_{\mathrm{sk}}=\frac{1}{4\pi}\iint_{\Omega}\mathbf{n}\cdot\left(\frac{\partial\mathbf{n}}{\partial u_x}\times\frac{\partial\mathbf{n}}{\partial u_y}\right)\mathrm{d}u_x\,\mathrm{d}u_y. \tag{12}$$

where $\mathbf{n}=(S_1, S_2, S_3)/S_0$ is the normalized Stokes vector, $u_x=k_x/k_0$ and $u_y=k_y/k_0$ are the direction cosines, and $\Omega$ is the observation region. The target value of $N_{\mathrm{sk}}$ given by Eq. (11) is denoted as $N_{\mathrm{sk}}^{\mathrm{target}}$, and the synthesis error is defined as $\Delta N_{\mathrm{sk}}=|N_{\mathrm{sk}}-N_{\mathrm{sk}}^{\mathrm{target}}|$.[44]

Secondly, the skyrmion type is evaluated by the Néel-like and Bloch-like components. Specifically, the radial and azimuthal projections of the unit in-plane Stokes vector are first averaged over the observation region, and the averaged vector is then normalized. The absolute values of the radial and azimuthal components of this normalized vector are defined as the Néel-like and Bloch-like components, respectively. With this definition, a Néel-type skyrmion ($\chi$=0 or π) has a Néel-like component of 1, and a Bloch-type skyrmion ($\chi$=±π/2) has a Bloch-like component of 1.

In Figure 4a, the Néel-like and Bloch-like components of both states alternate with a period of π, because the phase bias $\Delta\delta$ rotates the in-plane Stokes vector by $-\Delta\delta$, as derived in Section 2.3. By fitting the Néel-like and Bloch-like components near each maximum with |cos(·)| and |sin(·)| functions of $\Delta\delta$, the Néel-type states are found at $\Delta\delta=0.474\pi$ and $1.406\pi$ for target $N_{\mathrm{sk}}=-1$ (solid lines) and at $\Delta\delta=0.440\pi$ and $1.442\pi$ for target $N_{\mathrm{sk}}=+1$ (dashed lines). The Bloch-type states are found at $\Delta\delta=0.939\pi$ and $1.942\pi$ for target $N_{\mathrm{sk}}=-1$ and at $\Delta\delta=0.944\pi$ and $1.937\pi$ for target $N_{\mathrm{sk}}=+1$. According to the conditions in Section 2.3, these positions are determined by $\delta_{\mathrm{MRR}}$, and $\delta_{\mathrm{MRR}}\approx-0.941\pi$ is obtained from the far fields in Section 3.1 over the observation region. With this $\delta_{\mathrm{MRR}}$, the theoretical values are $\Delta\delta=0.441\pi$ and $1.441\pi$ for the Néel type and $\Delta\delta=0.941\pi$ and $1.941\pi$ for the Bloch type in both states. The differences in $\Delta\delta$ between the simulated positions and these theoretical values are less than 0.04π for target $N_{\mathrm{sk}}=-1$ and less than 0.005π for target $N_{\mathrm{sk}}=+1$.

Meanwhile, the actual $N_{\mathrm{sk}}$ realized by our SIGMA changes by less than 0.015 over the whole sweep for both states. For target $N_{\mathrm{sk}}=-1$, Figure 4b–d compare the Néel-type and Bloch-type skyrmions generated at $\Delta\delta$=1.385π and 1.949π, respectively, and the corresponding results for target $N_{\mathrm{sk}}=+1$ are given in Section S1 of the Supporting Information. Between these two states, the in-plane Stokes vector in Figure 4c changes from the radial direction to the azimuthal direction, while the far-field intensity and the distribution of the normalized Stokes vectors on the Poincaré sphere remain almost unchanged. This shows that the phase bias $\Delta\delta$ tunes the skyrmion type independently of the skyrmion number and the far-field intensity.

In contrast, programming $l_{\mathrm{sub}}=1$ on the IG-MRRs gives $\Delta l=1$ and generates an anti-skyrmion, whose texture is only rotated by tuning $\Delta\delta$. As shown in Section S1 of the Supporting Information, the length of the averaged vector used to normalize the Néel-like and Bloch-like components stays below 0.15 over the whole sweep, whereas it is between 0.87 and 0.93 for the target $N_{sk}$=−1 state in Figure 4a. Therefore, the anti-skyrmion has no Néel-type or Bloch-type state.

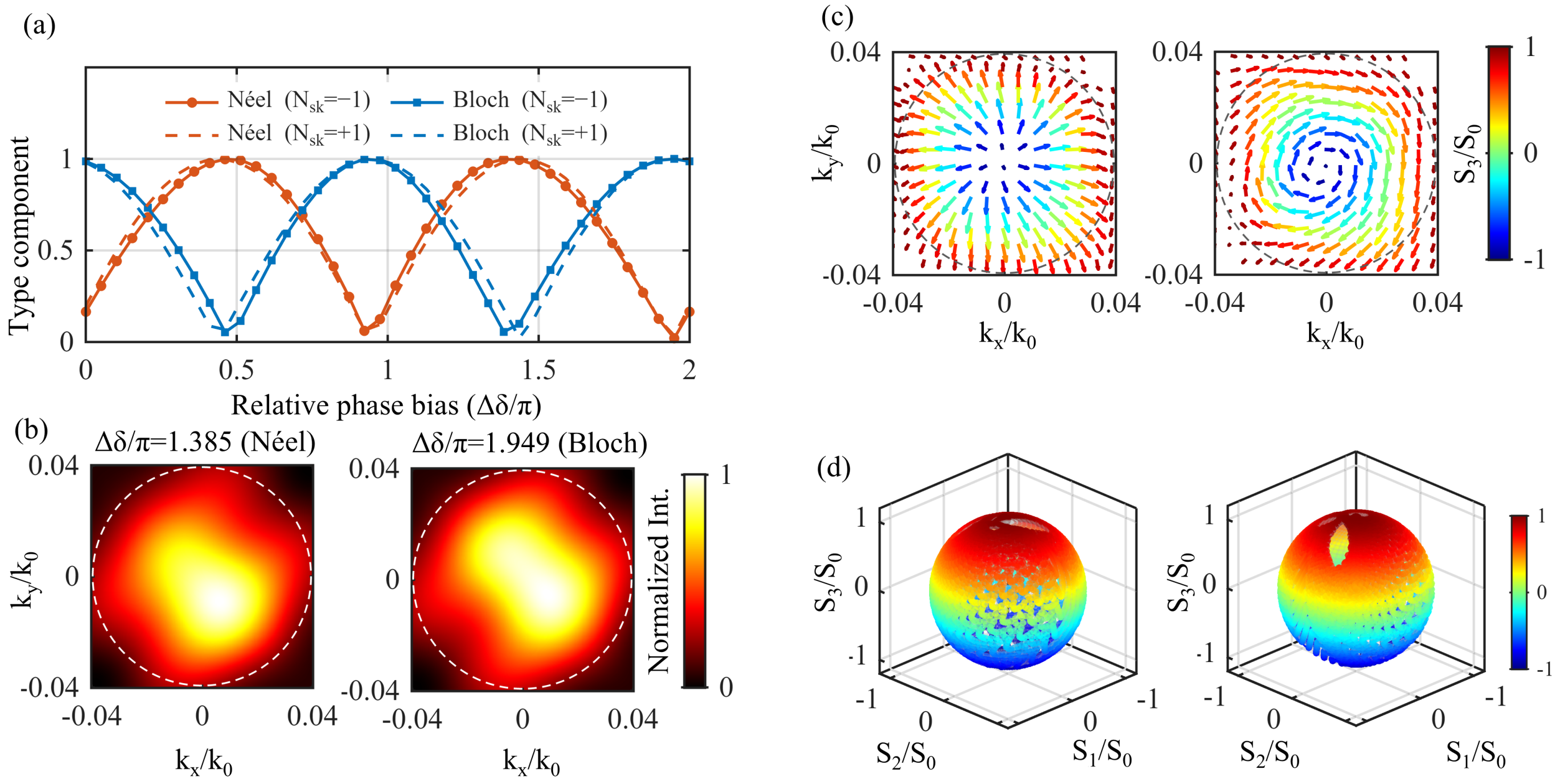

Figure 4. Phase-controlled switching between Néel-type and Bloch-type optical skyrmions in the 4×4 OPA with $\Delta l$=−1, for target $N_{sk}$=−1 and target $N_{sk}$=+1. (a) Néel-like component (orange) and Bloch-like component (blue) as functions of the phase bias $\Delta\delta$. Solid lines: target $N_{sk}$=−1 ($l_{sub}$=−1 on the IG-MRRs); dashed lines: target $N_{sk}$=+1 ($l_{sub}$=1 on the OG-MRRs). $\Delta\delta$ is swept from 0 to 2π with a step of 2π/39. The Néel-like and Bloch-like components are calculated as described in Section 3.2. The Néel-type and Bloch-type states of target $N_{sk}$=−1 shown in (b)–(d) are selected at $\Delta\delta$=1.385π and 1.949π, with a Néel-like component of 0.998 and a Bloch-like component of 1.000, respectively. Over the whole sweep, the actual $N_{sk}$ realized by our SIGMA changes by less than 0.015 for both states. (b) Normalized far-field intensity distributions of the selected states in the ($k_x/k_0$, $k_y/k_0$) plane. (c) Polarization maps of the selected states, where the arrows indicate the in-plane vector ($S_1/S_0$, $S_2/S_0$) and the color represents $S_3/S_0$. Dashed circles in (b) and (c) mark the observation region NA≤0.04. (d) Mapping of the normalized Stokes vector **n** onto the Poincaré sphere for the selected states, where the color also represents $S_3/S_0$. (b)–(d) show the target $N_{sk}$=−1 states as an example, and the corresponding results for target $N_{sk}$=+1 are given in Section S1 of the Supporting Information.

Overall, the skyrmions with target $N_{sk}$=±1 are generated by programming $l_{sub}$=−1 on the IG-MRRs or $l_{sub}$=1 on the OG-MRRs, both of which give $\Delta l$=−1. For these skyrmions, increasing $\Delta\delta$ from about 0.44π to 0.94π switches the Néel type to the Bloch type, and the two types alternate every π/2, while the actual $N_{sk}$ realized by our SIGMA remains almost unchanged. When $l_{sub}$=1 is programmed on the IG-MRRs instead, $\Delta l$=1 and SIGMA generates an anti-skyrmion, whose type cannot be switched by $\Delta\delta$.

### 3.3. Dynamic Reconfiguration of the Skyrmion Number

With the phase bias fixed at $\Delta\delta$=0, the skyrmion number $N_{sk}$ can be controlled by changing the azimuthal phase order $l_{sub}$. The same 4×4 SIGMA is used here. Specifically, we program $l_{sub}$=−1 and −2 on the IG-MRR subarray and $l_{sub}$=1 and 2 on the OG-MRR subarray for the target states $N_{sk}$=−1, −2, 1, and 2, respectively, as shown in Figure 5a. The input phase delay of every MRR in the other subarray is set to $\Phi_{pq}$=0. The amplitude ratio $a_{IG}/a_{OG}$ is set to 1.52 for the negative states and 1.02 for the positive states, as in Section 3.2. Therefore, switching among the four target states requires only changes in $l_{sub}$ and $a_{IG}/a_{OG}$.

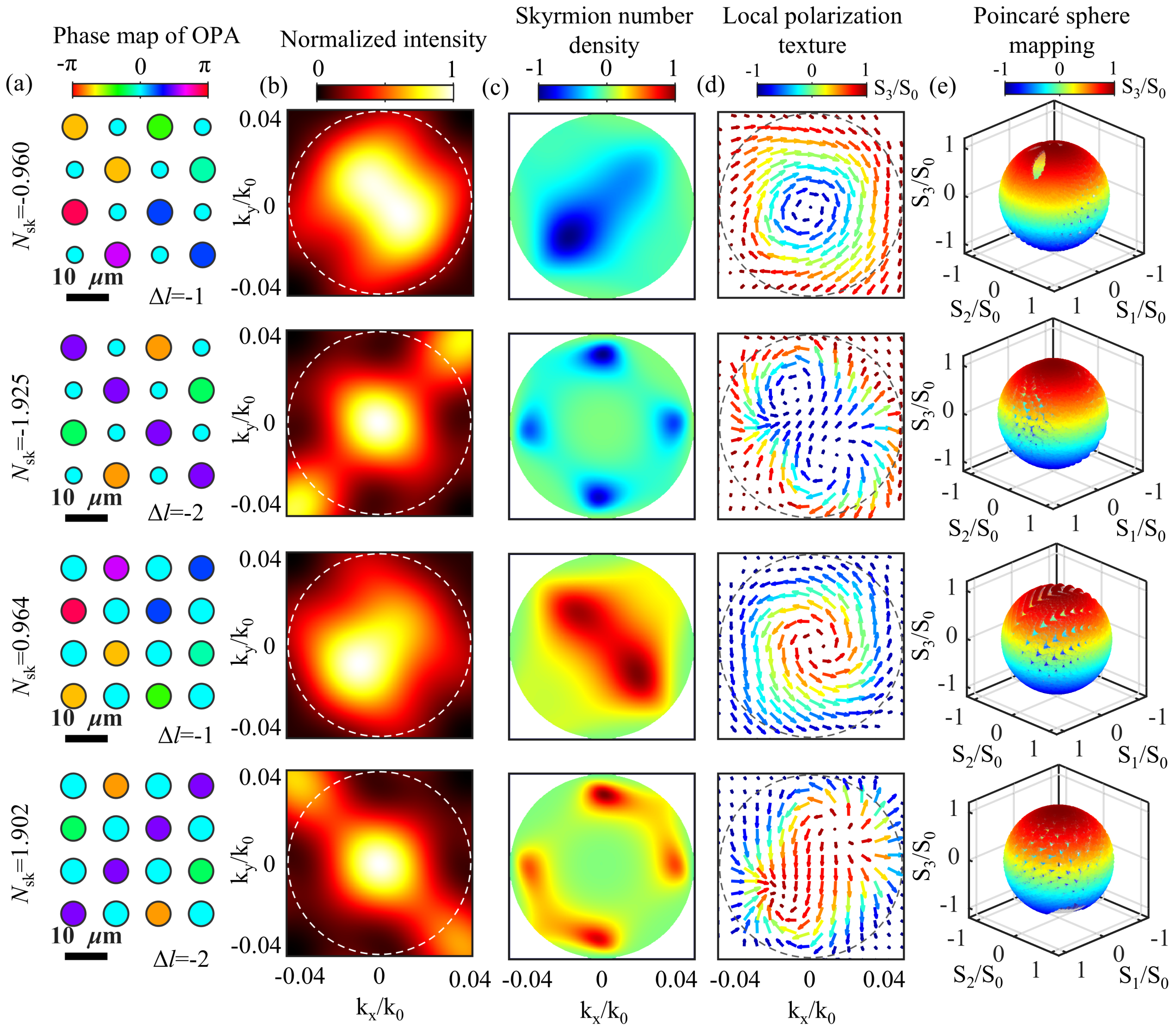

Figure 5. Dynamic reconfiguration of the optical skyrmion number by the programmable azimuthal phase of the OPA. Each row corresponds to one skyrmion state, and the columns show (a) the phase map of the OPA, (b) the normalized far-field intensity, (c) the skyrmion number density, normalized by its maximum absolute value within the observation region for each state, (d) the local polarization texture, and (e) the corresponding Poincaré-sphere mapping of the normalized Stokes vector. From top to bottom, the four states correspond to $l_{sub}$=−1 and −2 programmed on the IG-MRRs and $l_{sub}$=1 and 2 programmed on the OG-MRRs ($\Delta l$=−1, −2, −1, and −2), with actual $N_{sk}$ values of −0.960, −1.925, 0.964, and 1.902, respectively. In (d), the arrows indicate the in-plane vector ($S_1/S_0$, $S_2/S_0$). In (d) and (e), the colors represent $S_3/S_0$. In (a), the marker area is proportional to the input power of each MRR. Dashed circles in (b) and (d) mark the observation region NA≤0.04.

With these settings, the generated states are shown in Figure 5, and the actual $N_{sk}$ realized by our SIGMA is −0.960, −1.925, 0.964, and 1.902 for $l_{sub}$=−1, −2, 1, and 2, respectively. These results agree with the target values given by Eq. (11) in both sign and magnitude, with $\Delta N_{sk}$ below 0.1 for all four states.

The far-field intensity is concentrated near the beam center for $|l_{sub}|$=1, whereas two additional bright regions appear along a diagonal for $|l_{sub}|$=2, as shown in Figure 5b. Here, the eight MRRs in the programmed subarray sample the azimuthal phase $l_{sub}\varphi_{pq}$ at their azimuthal positions $\varphi_{pq}$. For $|l_{sub}|$=2, this discrete sampling produces the two bright regions, because the positions of these MRRs are not symmetric under a 90° rotation about the array center.

The textures in Figures 5c–e are consistent with these skyrmion numbers, namely the skyrmion number density has the same sign as the target $N_{sk}$, the in-plane Stokes vector winds around the beam center $|l_{sub}|$ times, and the normalized Stokes vectors **n** cover almost the entire Poincaré sphere. Together, these results confirm that SIGMA can dynamically reconfigure the skyrmion number in both sign and magnitude only by the phase shifters, and the four generated states form the four-symbol set used in Section 4.

### 3.4. Extension of the Programmable Range of the Skyrmion Number

The programmable range of $N_{sk}$ is further extended by increasing the array size. The simulated $N\times N$ arrays have $N$=4, 6, 8, 10, and 12, each containing $N^2/2$ IG-MRRs and $N^2/2$ OG-MRRs in a checkerboard arrangement. All arrays use the same pitch and the same observation region as the 4×4 SIGMA in Section 3.2. Following the phase configurations in Section 3.3, we set $l_{sub}=N_{sk}^{target}$ on the OG-MRR subarray for positive target values and on the IG-MRR subarray for negative target values. No azimuthal phase is applied to the other subarray.

Unlike the 4×4 SIGMA in Sections 3.2 and 3.3, which uses all 16 MRRs and the amplitude ratios set in Section 3.2, the input settings here are optimized for each array size $N$ and target value $N_{sk}^{target}$. Specifically, the input amplitudes of the IG-MRRs and OG-MRRs outside two separately selected radii from the array center are set to zero, and the remaining MRRs are denoted as the active MRRs. The two radii, the amplitude ratio $a_{IG}/a_{OG}$, and the phase bias $\Delta\delta$ are optimized to minimize $\Delta N_{sk}$. The phase bias $\Delta\delta$ is included in the optimization because the FDTD fields retain the residual circular polarization components neglected in the ideal model in Section 2.3. Because a high-order texture can accumulate $N_{sk}$ in regions of low intensity, $N_{sk}$ is also calculated only over the pixels with at least 3% of the maximum intensity, as for the received fields in Section S4 of the Supporting Information, and $\Delta N_{sk}$ in this section denotes the larger deviation of the two $N_{sk}$ values from $N_{sk}^{target}$. The search ranges and the settings of all synthesized states are given in Section S2 of the Supporting Information.

We adopt an error tolerance of $\Delta N_{sk}\leq 0.35$ as the synthesis criterion in this study. Figure 6a presents the results using a normalized synthesis score, which equals 1 for $\Delta N_{sk}\leq 0.15$, decreases linearly to 0 as $\Delta N_{sk}$ increases from 0.15 to 0.80, and remains 0 for larger errors.

Among the evaluated targets, the largest values that satisfy the synthesis criterion $\Delta N_{sk}\leq 0.35$ are 2, 4, 5, 7, and 8 for the 4×4, 6×6, 8×8, 10×10, and 12×12 arrays, respectively, as shown in Figure 6a. Up to the largest target value of each array, the actual $N_{sk}$ stays close to $N_{sk}^{target}$, while for larger target values, it falls below $N_{sk}^{target}$, as shown in Figure 6b. For the largest target values that satisfy the criterion, the selected settings extend the active MRRs of the programmed subarray to a larger radius and keep only the central MRRs of the other subarray active, because a small aperture gives a broad central lobe without sidelobes in the observation region, whereas the vortex needs many MRRs. Larger arrays provide more MRRs to sample the programmed azimuthal phase $l_{sub}\varphi_{pq}$, allowing the synthesis of larger target skyrmion numbers $N_{sk}^{target}$.

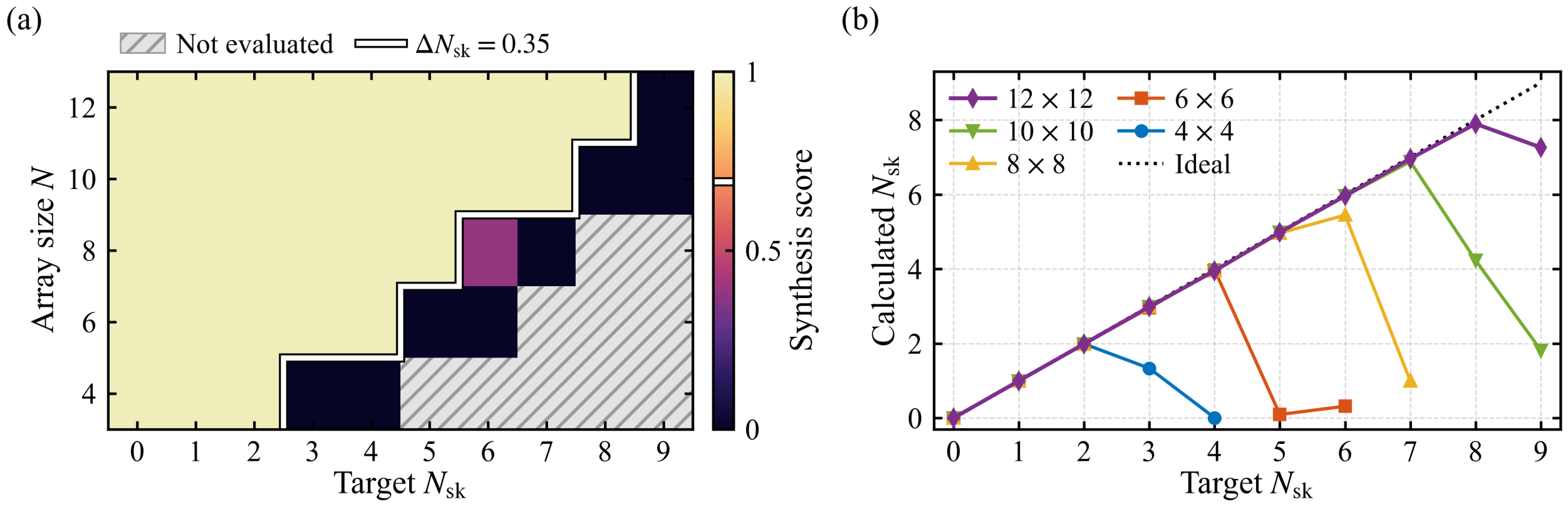

Figure 6. Extension of the programmable range of the skyrmion number with the array size. For each array size and each target skyrmion number, the active MRRs of each subarray, the amplitude ratio, and the phase bias are optimized as described in Section 3.4. (a) Synthesis score for different array sizes $N$ and target skyrmion numbers. The synthesis criterion is $\Delta N_{sk} \leq 0.35$. The white line marks this criterion. Hatched gray cells are targets that are not evaluated, because the search stops after two consecutive failures (Section S2 of the Supporting Information). (b) Actual $N_{sk}$ as a function of the target $N_{sk}$ for 4×4, 6×6, 8×8, 10×10, and 12×12 OPAs. For each state, the $N_{sk}$ value that deviates more from the target value is shown. The black dotted line represents the ideal relation $N_{sk}=N_{sk}^{target}$. The largest target skyrmion number that satisfies the synthesis criterion increases from 2 for the 4×4 array to 8 for the 12×12 array, which is consistent with the achievable skyrmion number being limited by the number of MRRs in the subarray that samples the azimuthal phase.

For the targets that satisfy the synthesis criterion, the negative states give actual $|N_{sk}|$ values within 0.01 of those of the corresponding positive states, and the textures of the 12×12 SIGMA for $N_{sk}^{target}$ from 1 to 8 and −8 are shown in Section S2 of the Supporting Information. Therefore, SIGMA can extend the programmable range of $N_{sk}$ by increasing the array size without redesigning the MRR emitters, and the 12×12 SIGMA covers target values $N_{sk}^{target}$ from −8 to 8.

### 3.5. Tolerance to Fabrication Variations

In silicon photonic chips, fabrication errors often make the devices deviate from their design and degrade their performance.[45] Previous studies have reported that variations in waveguide geometries affect the resonance wavelengths of fabricated on-chip devices.[45,46] We therefore evaluate the effects of fabrication errors on the skyrmion synthesis of SIGMA and investigate calibration methods that recover the target states only by the laser and the modulation network.

Here, we consider three types of fabrication errors. The first type includes errors in the ring waveguide width $W_{wg}$, bus waveguide width $W_{bus}$, waveguide height $H_{wg}$, coupling gap $s$, grating length $l_g$, grating width $w_g$, and grating etch depth $d_g$. The grating etch depth $d_g$ differs between the designed IG- and OG-MRRs, so the gratings require two shallow-etch steps. These geometric errors change the resonance wavelengths, loaded quality factors, and upward emission efficiencies of the MRRs, as described in Section S3.A of the Supporting Information. The second type is the local fabrication differences between individual MRRs, which shift the resonance wavelength of each MRR independently.[45,46] The third type is the overlay errors between the grating and waveguide layers, which displace the gratings relative to the ring waveguides[47] and thus change the emitted far fields.

For the geometric errors, we use the MRR model described in Section 3.1 to calculate the optical responses of the IG- and OG-MRRs. These errors are then sampled in a Monte Carlo analysis. In each trial, the same sampled deviation of each geometric parameter is applied to all MRRs, except for $d_g$. The deviation of $d_g$ is sampled independently for the two subarrays and shared by all MRRs within each subarray to represent the two shallow-etch steps. For the local fabrication differences, we add an independent random resonance shift to each MRR. These shifts follow a zero-mean Gaussian distribution with a standard deviation of $\sigma_{in}$. We further evaluate the effects of far-field changes by separately substituting the single-MRR far fields calculated with geometric and overlay errors into SIGMA. The fabrication error settings are selected with reference to previously published studies,[45,46,48–51] and their values are given in Section S3.C of the Supporting Information.

The calibration methods use only the laser and the modulation network of SIGMA. The effects shared by the MRRs in each subarray are reduced by tuning the laser wavelength and adjusting $\Delta\delta$ and $a_{IG}/a_{OG}$, while the effects of the independent resonance shifts are reduced by the calibration of the input phase of each MRR (phase calibration) and, when needed, its input amplitude (amplitude and phase calibration). Both calibrations assume that the complex response of each MRR at the operating wavelength is known.

Table 2. Effect of Fabrication Errors on the 4×4 SIGMA and Their Calibration

| Fabrication errors | Calibration | Pass | $T$ | $P/P_0$ |
|---|---|---|---|---|
| Waveguide width and height | Laser tuning | 100% | 0.984 | 0.99 |
| Etch depth $d_g$, 5 nm | Laser tuning, $\Delta\delta$ adjustment | 100% | 0.999 | 0.61 |
| Combined geometric | Laser tuning, $\Delta\delta$ adjustment | 100% | 0.999 | 0.59 |
| Combined geometric and $\sigma_{in}$=0.5 nm | Laser tuning, phase calibration, $a_{IG}/a_{OG}$ adjustment | 94.8% | 0.999 | 0.51 |
| Combined geometric and $\sigma_{in}$=0.5–1.5 nm | Laser tuning, amplitude and phase calibration | 99.9% | 0.997 | 0.56–0.36 |
| Overlay, 20–30 nm | $\Delta\delta$ and $a_{IG}/a_{OG}$ adjustment | 100% | 0.999 | 0.92–0.80 |

Note: Pass denotes the lowest fraction of trials satisfying $\Delta N_{sk}\leq 0.35$ among the four target states in Figure 5. $T$ is the lowest mean target type component among the four Néel-type and Bloch-type states with target $N_{sk}$=±1 (Section 3.2), and $P/P_0$ is the mean emitted power relative to the nominal design. Combined geometric cases use a standard deviation of 5 nm in $d_g$. The overlay row lists the worst fixed displacement. Detailed settings are provided in Section S3 of the Supporting Information.

With these calibration methods, SIGMA recovers the target states under all the evaluated fabrication errors, as summarized in Table 2. The geometric errors are shared by the MRRs of each subarray, so laser tuning and the adjustment of $\Delta\delta$ keep all four target states in Figure 5 within the synthesis criterion in every trial and restore the lowest mean target type component to 0.999. When the independent resonance shifts with $\sigma_{in}$=0.5 nm are also included, the phase calibration together with the adjustment of $a_{IG}/a_{OG}$ gives a lowest pass rate of 94.8% among the four target states, and the amplitude and phase calibration raises it to at least 99.9% at all the tested values of $\sigma_{in}$. Under the overlay errors, the adjustment of $\Delta\delta$ and $a_{IG}/a_{OG}$ also recovers the target types. Meanwhile, the emitted power decreases to different extents, as listed in Table 2. Therefore, only by laser tuning and the phase shifters in the modulation network, SIGMA can still generate the target skyrmion numbers and skyrmion types under the evaluated fabrication errors, which shows that the proposed

SIGMA is robust to fabrication errors. Section S3 of the Supporting Information provides the trial numbers, the residual calibration errors, and the detailed comparisons.

## 4. SKYRMION-ENCODED FREE-SPACE OPTICAL COMMUNICATION

In this section, the skyrmion textures generated by SIGMA are further evaluated as information carriers in a four-symbol free-space optical communication channel. The symbol error rate (SER) of the skyrmion channel, which decodes each symbol from $N_{sk}$, is compared with that of a conventional LG-OAM channel.

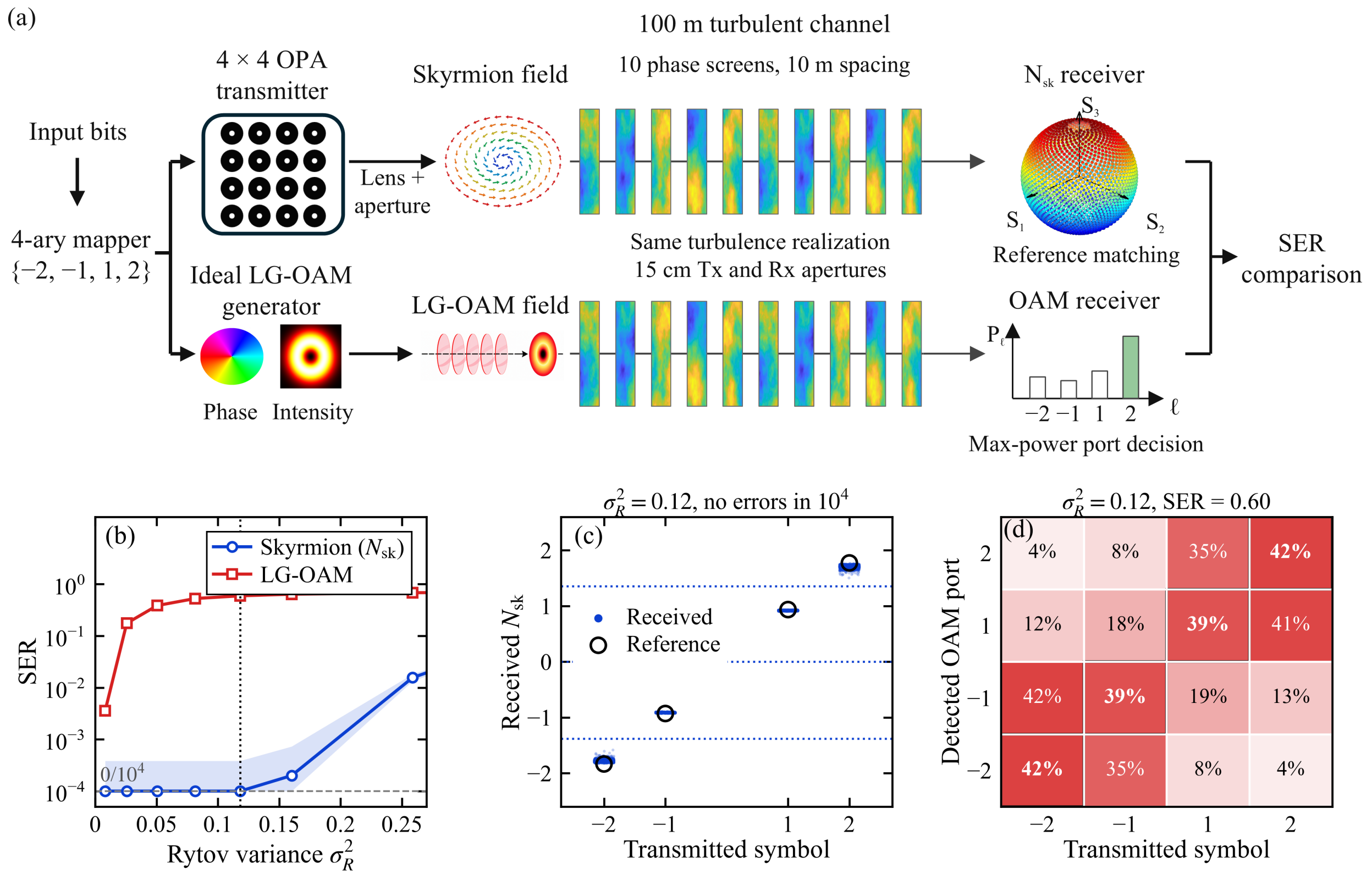

Figure 7. Skyrmion-encoded free-space optical communication through atmospheric turbulence. (a) Schematic of the skyrmion channel and the LG-OAM reference channel with the same four-symbol set {−2, −1, 1, 2}. Both channels use a 15 cm transmitting aperture and a 15 cm receiving aperture, and propagate over 100 m through ten phase screens. (b) SER versus the plane-wave Rytov variance $\sigma_R{}^2$ for the skyrmion channel (blue circles) and the LG-OAM channel (red squares), from $10^4$ paired trials at each turbulence strength. Shaded bands show 95% Wilson confidence intervals. Points without observed errors are plotted on the gray dashed line, which marks the one-error level of $10^{-4}$. The vertical dotted line marks $\sigma_R{}^2$=0.12, which is the turbulence strength used in (c) and (d). (c) Received $N_{sk}$ values of the $10^4$ trials at $\sigma_R{}^2$=0.12. Black circles indicate the references obtained without turbulence, and blue dotted lines mark the decision thresholds. (d) Fraction of the trials of each transmitted symbol detected at each OAM port, from the same $10^4$ trials. Boxed cells mark the transmitted orders.

The setup of the two channels is illustrated in Figure 7a. In the skyrmion channel, the transmitter uses the four skyrmion states in Figure 5 to encode the symbols {−2, −1, 1, 2}. To launch the skyrmion beam into the free-space channel, the diverging beam emitted by SIGMA is collimated by an ideal lens with a focal length of $f$. This lens maps each emission direction ($k_x/k_0$, $k_y/k_0$) onto the transverse position $f\cdot(k_x/k_0, k_y/k_0)$ at its back focal plane without changing the polarization. A circular aperture with a radius of $0.04f$ is placed at the back focal plane of the lens, so that only the emission within the observation region of NA≤0.04 is transmitted. Here, $f$ is chosen so that the root-mean-square (RMS) radius of the four skyrmion beams is approximately 3 cm, as given in Section S4 of the Supporting Information. For the LG-OAM reference channel, uniformly polarized LG modes with OAM orders $l$=−2, −1, 1, and 2 are used to represent the same four symbols. To compare the two channels under the same conditions, each LG beam has an RMS radius of 3 cm and the same power as the skyrmion beams, and both channels use a transmitting aperture with a diameter of 15 cm.

The turbulent channel is modeled by the split-step propagation method with multiple random phase screens.[52,53] Here, the channel length is set to $L$=100 m,[54] and ten phase screens are placed at equal intervals along the channel. Each phase screen is generated from the modified von Kármán power spectrum with an inner scale of 5 mm and an outer scale of 20 m, and its low-frequency components are compensated by subharmonics.[55] The turbulence strength is described by the refractive-index structure constant $C_n^2$, and the corresponding plane-wave Rytov variance is given by $\sigma_R^2=1.23C_n^2k_0^{7/6}L^{11/6}$.[56] Along the channel, the integrated turbulence strength $C_n^2L$ is divided equally among the ten screens. In each Monte Carlo trial, the transmitted symbol is selected uniformly from the four-symbol set, and the beams of the two channels propagate through the same set of phase screens. Seven values of the Rytov variance $\sigma_R^2$ between 0.008 and 0.26 are evaluated, and the SER at each value is estimated from $10^4$ trials.

At the receiver of each channel, the beam is collected by a circular aperture with a diameter of 15 cm, which captures more than 95% of the power of every transmitted beam after the propagation without turbulence. For the skyrmion channel, we calculate $N_{sk}$ of the received field by applying the integral in Eq. (12) in the transverse plane.[57] The integral is evaluated on the selected pixels within the receiving aperture, and the pixel selection and the numerical evaluation of $N_{sk}$ are described in Section S4 of the Supporting Information. The calculated $N_{sk}$ is then assigned to the closest of the four reference values, which are obtained by propagating the four skyrmion states through the same channel without turbulence. For the LG-OAM channel, the received field is projected onto the four LG modes that have propagated over the same 100 m distance without turbulence, which corresponds to a mode demultiplexer with four output ports.[58,59] The symbol is then decided by the output port with the largest power.

Figure 7b compares the SERs of the two channels as functions of $\sigma_R^2$. For the LG-OAM channel, the SER already reaches 0.18 at $\sigma_R^2$=0.026 and 0.60 at $\sigma_R^2$=0.12. In contrast, no symbol errors are observed for the skyrmion channel in $10^4$ trials up to $\sigma_R^2$=0.12. At $\sigma_R^2$=0.16, the SER of the skyrmion channel is $2\times10^{-4}$, which is about three orders of magnitude lower than the SER of 0.64 for the LG-OAM channel. Even at $\sigma_R^2$=0.26, the SER of the skyrmion channel is 0.016, which is still about 40 times lower than that of the LG-OAM channel.

This difference can be seen from the received $N_{sk}$ values and the detected OAM ports at $\sigma_R^2$=0.12, as shown in Figures 7c and 7d, respectively. In Figure 7c, all $10^4$ received $N_{sk}$ values remain close to the reference values and stay within the decision regions. The standard deviation of the received $N_{sk}$ is below 0.04 for every symbol. In contrast, Figure 7d shows that the trials of the LG-OAM channel are detected at all four output ports, and only 39%–42% of the trials are detected at the port of the transmitted symbol. Most of the errors occur between the two OAM orders with the same sign, such as $l$=−2 and $l$=−1. This is because the phase distortion directly changes the azimuthal phase of the received field and mainly transfers power between neighboring OAM orders.

The skyrmion channel is more robust because the phase distortion is the same for the two circularly polarized components of the skyrmion beam. Thus, the relative phase between these two components, which determines the local polarization, is not changed. Meanwhile, the diffraction between the phase screens only deforms the polarization texture continuously. Therefore, the topological invariant $N_{sk}$ remains nearly unchanged.[11,54,60] These results show that SIGMA can directly serve as a programmable transmitter for skyrmion-encoded free-space optical communication, in which the four symbols are switched only by the phase shifters.

## 5. CONCLUSION

In this paper, we propose and numerically demonstrate SIGMA, a silicon optical phased array that uses two types of chiral microring emitters to generate and program optical skyrmions. The optimized IG- and OG-MRRs emit LCP and RCP light, with circular polarization fractions of 88.08% and 93.68%, respectively. Configuring the phase shifters enables the 4×4 array to generate four target states with actual $N_{sk}$ values ranging from −1.925 to 1.902 and continuously switch the states with target $N_{sk}=\pm1$ between the Néel and Bloch types. Moreover, enlarging the array to 12×12 extends the range of target $N_{sk}$ from −8 to 8 without redesigning the MRR emitters, so the programmable range of SIGMA grows with the array size. Under the simulated fabrication errors, the target skyrmion numbers and types are recovered only by laser tuning and the phase shifters in the modulation network, which shows that SIGMA is robust to fabrication errors without any change to the fabricated MRRs. In a four-symbol communication link through 100 m of simulated atmospheric turbulence, decoding based on $N_{sk}$ shows no symbol errors in $10^4$ trials up to a Rytov variance of 0.12, whereas the SER of the LG-OAM channel already reaches 0.60. This indicates that the skyrmion states programmed by SIGMA keep their skyrmion numbers through turbulence and can serve as robust information carriers. These numerical results demonstrate the potential of SIGMA as a compact, programmable skyrmion transmitter for turbulence-tolerant free-space optical communication.

## ASSOCIATED CONTENT

### Data Availability Statement

Data underlying the results presented in this paper are not publicly available at this time but may be obtained from the authors upon reasonable request.

### Supporting Information

The Supporting Information is available free of charge. Skyrmion-type switching for target $N_{sk}$=+1 and the anti-skyrmion; search settings and synthesized states for different array sizes; fabrication tolerance; transmitter and receiver of the skyrmion channel (PDF)

## AUTHOR INFORMATION

### Corresponding Author

Tian Zhang – State Key Laboratory of Information Photonics and Optical Communications, Beijing University of Posts and Telecommunications, Beijing 100876, China; Email: ztian@bupt.edu.cn

### Author Contributions

**Zili Cai:** Conceptualization, Methodology, Software, Validation, Formal analysis, Investigation, Data curation, Visualization, Writing – original draft, Writing – review & editing. **Tian Zhang:**

Supervision, Writing – review & editing, Resources, Funding acquisition. **Qi Chen:** Writing – review & editing. **Jubo Hao:** Validation, Writing – review & editing. **Jian Dai:** Project administration. **Kun Xu:** Project administration, Funding acquisition, Resources.

**Notes**

The authors declare no competing financial interest.

## ACKNOWLEDGMENTS

This work was supported by the National Natural Science Foundation of China (62171055, 62135009, 62471062); Fundamental Research Funds for the Central Universities (ZDYY202102); Super Computing Platform of Beijing University of Posts and Telecommunications.

The authors used Claude (Anthropic) and ChatGPT (OpenAI) to improve the grammar and readability of the manuscript and the Supporting Information, and reviewed and verified all edited text.

# Supporting Information

## Dynamically Reconfigurable Optical Skyrmions Enabled by a Silicon Microring Optical Phased Array for Robust Free-Space Communication

Zili Cai, Tian Zhang, Qi Chen, Jubo Hao, Jian Dai, and Kun Xu

### S1. Skyrmion-Type Switching at $N_{sk} = +1$ and the Anti-Skyrmion at $\Delta l = +1$

Figure S1 shows the Néel-type and Bloch-type states with a target skyrmion number of +1, as discussed in Section 3.2. These states are generated by setting $l_{sub}$=1 on the OG-MRRs and $a_{IG}/a_{OG}$=1.02. The two states have nearly the same far-field intensity and cover nearly the same region of the Poincaré sphere. At each position, the angle between the in-plane Stokes vectors of the Néel-type and Bloch-type states is approximately $\pi/2$. The states with a target skyrmion number of −1 in Figure 4 show the same rotation.

The anti-skyrmion in Figure S2 is generated by setting $l_{sub}$=1 on the IG-MRRs ($\Delta l$=1) and using the amplitude weights of Figure 4. For this saddle-like texture, the angle $\chi$ between the in-plane Stokes vector and the local radial direction depends on the azimuthal position. Changing $\Delta\delta$ rotates the in-plane Stokes vectors at each position. The type components defined in Section 3.2 use the spatial average of ($\cos\chi$, $\sin\chi$) over pixels within NA≤0.04 where $(S_1, S_2)/S_0$ has a magnitude greater than 0.05. The averaged vector is then normalized by its length. For the anti-skyrmion, this length remains below 0.15 throughout the $\Delta\delta$ sweep, compared with 0.87–0.93 for the $\Delta l$=−1 states.

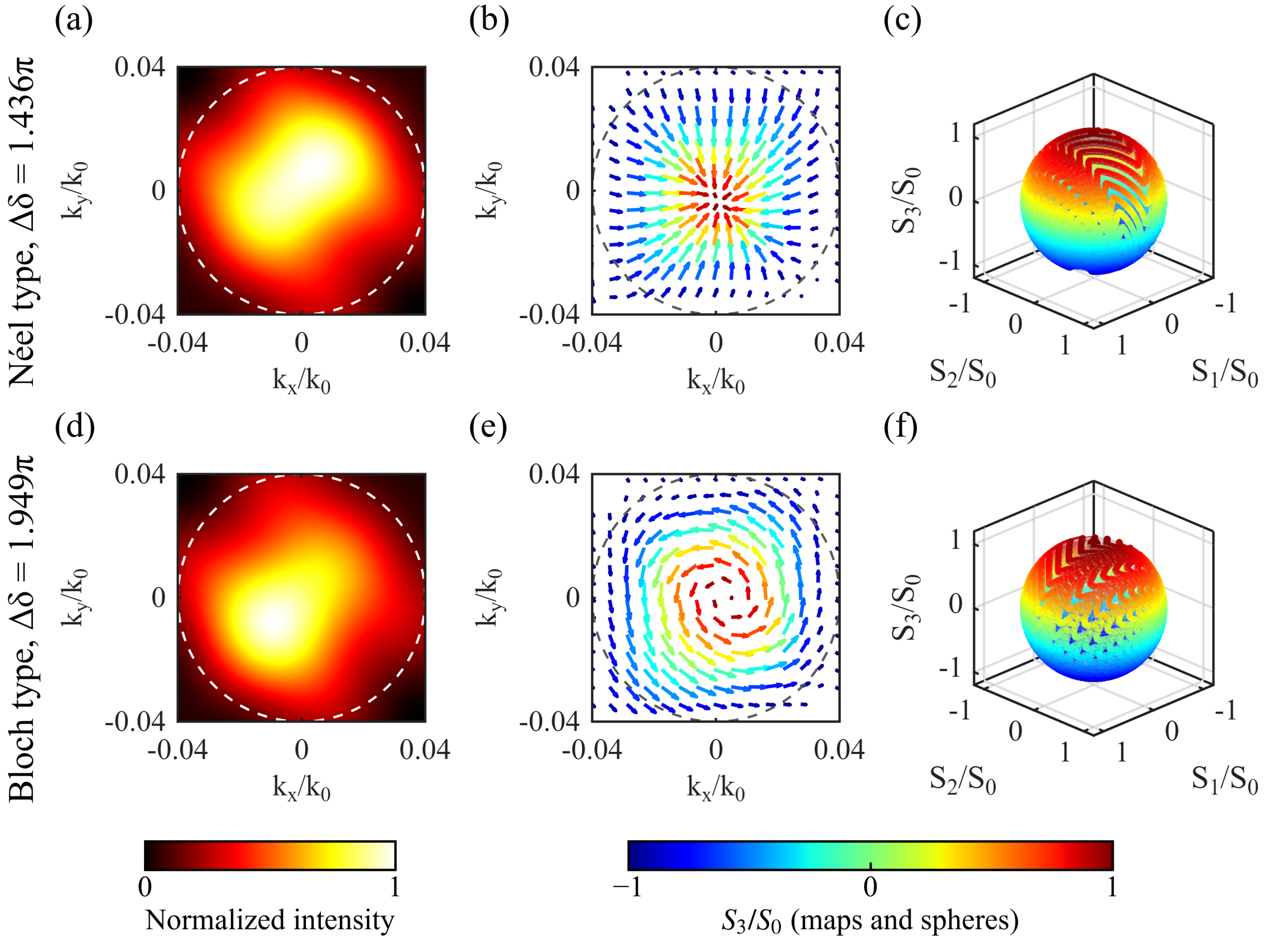

**Figure S1.** Néel-type and Bloch-type states with a target $N_{sk}$=+1 in the 4×4 SIGMA. The OG-MRRs use $l_{sub}$=1, with $a_{IG}/a_{OG}$=1.02. (a)–(c) Néel-type state at $\Delta\delta$=1.436π, with $N_{sk}$=0.966 and a Néel-like component of 1.000. (d)–(f) Bloch-type state at $\Delta\delta$=1.949π, with $N_{sk}$=0.964 and a Bloch-like component of 0.999. (a), (d) Normalized far-field intensity in the ($k_x/k_0$, $k_y/k_0$) plane. (b), (e) Polarization maps. Arrows show ($S_1/S_0$, $S_2/S_0$), and color shows $S_3/S_0$. (c), (f) Normalized Stokes vectors **n** mapped onto the Poincaré sphere. Dashed circles in (a), (b), (d), and (e) mark the observation region NA≤0.04.

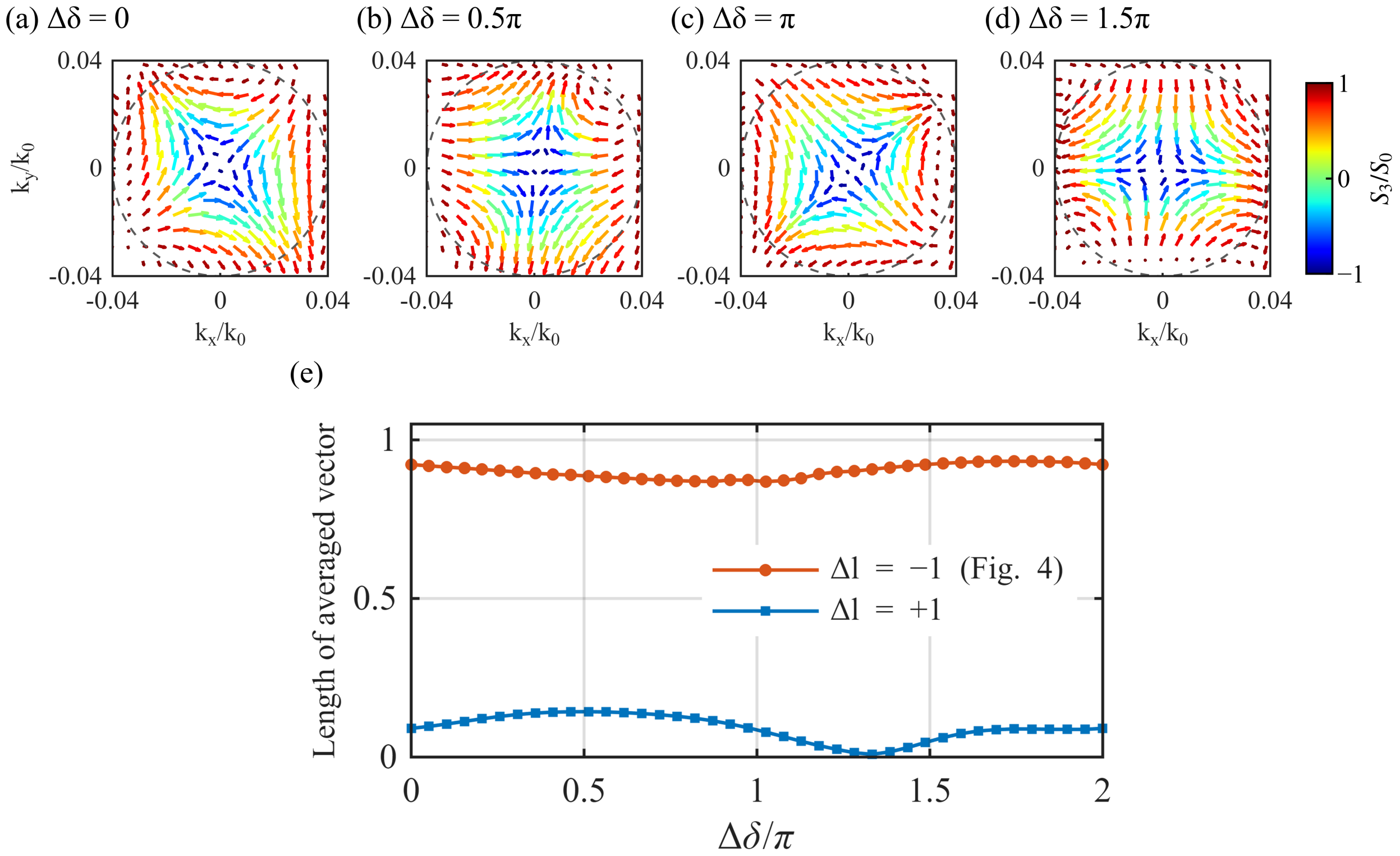

**Figure S2.** Anti-skyrmion in the 4×4 SIGMA with $l_{sub}$=1 on the IG-MRRs ($\Delta l$=1) and the amplitude weights of Figure 4. (a)–(d) Polarization maps at $\Delta\delta$=0, 0.5π, π, and 1.5π. Arrows show ($S_1/S_0$, $S_2/S_0$), and color shows $S_3/S_0$. Dashed circles mark the observation region NA≤0.04. (e) Length of the spatially averaged vector (cos $\chi$, sin $\chi$) used to normalize the type components, as a function of $\Delta\delta$. The curves compare $\Delta l$=−1 (Figure 4) and $\Delta l$=1.

## S2. Search Settings and Synthesized States for Different Array Sizes

The search in Section 3.4 selects the active MRRs, amplitude ratio, and phase bias for each array size $N$ and target skyrmion number $N_{sk}^{target}$. Only the IG-MRRs within the radius $R_{act,IG}$ and the OG-MRRs within the radius $R_{act,OG}$ of the array center are used. The search first uses a common radius $R_{act,IG}$=$R_{act,OG}$ from 10 μm to the array corner in steps of 2.5 μm, plus one value 0.1 μm beyond the corner. The amplitude ratio $a_{IG}/a_{OG}$ is selected from 20 values spaced evenly on a logarithmic scale. These values make the power emitted into NA≤0.04 by one MRR in the vortex subarray 0.1–15 times that of one MRR in the other subarray. The phase bias $\Delta\delta$ is selected from 36 values over 2π. Because $\Delta N_{sk}$≤0.15 already gives the maximum synthesis score of 1 in Figure 6a, a second search is made for targets with $\Delta N_{sk}$>0.15. It selects $R_{act,IG}$ and $R_{act,OG}$ separately from the same radius values, with $R_{act,IG}$≤$R_{act,OG}$, and further requires $R_{act,IG}$≤40 μm and $R_{act,OG}$≥40 μm for the 10×10 and 12×12 arrays, and $R_{act,IG}$≤40 μm and $R_{act,OG}$≥30 μm for the 8×8 array. It uses 25 amplitude ratios, which make the above power ratio 0.01–10, and 18 values of $\Delta\delta$ over 2π. The setting with the smaller $\Delta N_{sk}$ of the two searches is selected. Here, $\Delta N_{sk}$ is the larger deviation from $N_{sk}^{target}$ of $N_{sk}$ calculated by Eq. (12) and $N_{sk}$ calculated by the lattice method of Section S4 over the pixels with at least 3% of the maximum intensity within NA≤0.04. For the states in Table S1, recalculation on a 600×600 grid changes these two values by at most 0.032 and 0.005, respectively.

For positive targets, the vortex is on the OG-MRRs. The edge condition requires the mean $-S_3/S_0$ to be at least 0.8 in the outer region from 0.925 to 1 of the observation radius. The search first selects the setting with the smallest $\Delta N_{sk}$. If this setting does not meet the edge condition, the search selects the setting with the smallest $\Delta N_{sk}$ among those that meet both $\Delta N_{sk}$≤0.35 and the edge condition. For each array size, the search stops when two consecutive targets above 2 give $\Delta N_{sk}$>0.35. Table S1 lists the selected settings. For these states, the mean $-S_3/S_0$ at the edge is at least 0.93 for positive targets, and the mean $S_3/S_0$ at the edge is at least 0.91 for negative targets.

For negative targets, $l_{sub}$=$N_{sk}^{target}$ is applied to the IG-MRRs. The radii of the vortex subarray and the other subarray and the power ratio between them are kept the same as for the corresponding positive target, while $\Delta\delta$ is optimized separately. For the targets in Table S1, the resulting $|N_{sk}|$ differs from the positive-target value by at most 0.003 by Eq. (12) and 0.006 with the intensity mask.

To calculate the Poincaré-sphere coverage, the sphere is divided into 36×18 cells with equal intervals in azimuth and $S_3/S_0$. Coverage is the fraction of cells containing at least one mapped Stokes vector **n** from the observation region. For the states in Table S1, the coverage is 100% for positive targets and at least 99.8% for negative targets.

Figure S3 shows the polarization textures of the 12×12 SIGMA for the target values from 1 to 8 and −8. Within NA≤0.04, the intensity mask removes one pixel of the 300×300 grid for the target value 4 and no pixel for the other eight states, and the two $N_{\mathrm{sk}}$ values of each state differ by at most 0.04.

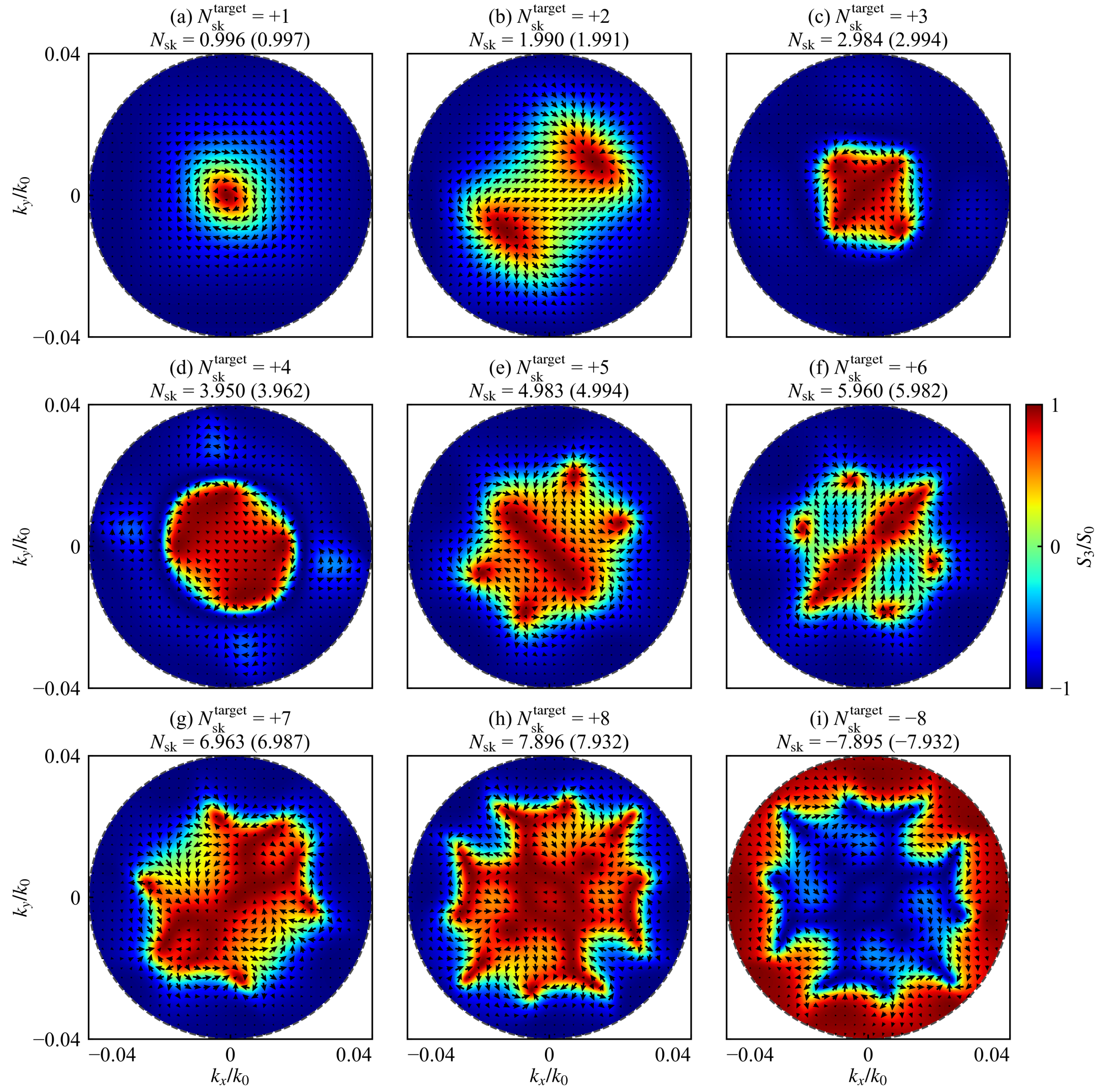

Figure S3. Polarization textures of the 12×12 SIGMA for $N_{\mathrm{sk}}^{\mathrm{target}}$ from 1 to 8 [(a)–(h)] and −8 [(i)], with the settings in Table S1. Colors show $S_3/S_0$, and arrows show ($S_1/S_0$, $S_2/S_0$). Dashed circles mark the observation region NA≤0.04. Each title gives $N_{\mathrm{sk}}$ calculated by Eq. (12) and, in parentheses, with the intensity mask.

TABLE S1

SETTINGS AND METRICS OF THE SYNTHESIZED STATES IN FIGURE 6

| $N$ | $N_{\mathrm{sk}}^{\mathrm{target}}$ | Active MRRs (IG/OG) | $a_{\mathrm{IG}}/a_{\mathrm{OG}}$ | $\Delta\delta/\pi$ | $N_{\mathrm{sk}}$ | $N_{\mathrm{sk}}$, masked | $\Delta\delta/\pi$ (negative target) | $N_{\mathrm{sk}}$ (negative target) | $N_{\mathrm{sk}}$, masked (negative target) |
|---|---|---|---|---|---|---|---|---|---|
| 4 | 1 | 8/8 | 0.321 | 0.944 | 0.996 | 0.997 | 0.833 | −0.996 | −0.997 |
| 4 | 2 | 8/8 | 0.321 | 1.278 | 1.990 | 1.991 | 1.944 | −1.988 | −1.989 |
| 6 | 1 | 8/8 | 0.321 | 0.944 | 0.996 | 0.997 | 0.833 | −0.996 | −0.997 |
| 6 | 2 | 8/8 | 0.321 | 1.278 | 1.990 | 1.991 | 1.944 | −1.988 | −1.989 |
| 6 | 3 | 16/16 | 0.321 | 0.111 | 2.959 | 2.963 | 1.333 | −2.959 | −2.963 |
| 6 | 4 | 8/18 | 0.606 | 1.222 | 3.975 | 3.980 | 0.333 | −3.974 | −3.980 |
| 8 | 1 | 8/8 | 0.321 | 0.944 | 0.996 | 0.997 | 0.833 | −0.996 | −0.997 |
| 8 | 2 | 8/8 | 0.321 | 1.278 | 1.990 | 1.991 | 1.944 | −1.988 | −1.989 |
| 8 | 3 | 30/30 | 0.321 | 0.056 | 2.984 | 2.994 | 0.833 | −2.983 | −2.992 |
| 8 | 4 | 30/30 | 0.477 | 1.222 | 3.950 | 3.962 | 1.278 | −3.948 | −3.956 |
| 8 | 5 | 8/32 | 1.436 | 0.000 | 4.963 | 4.974 | 1.889 | −4.964 | −4.976 |
| 10 | 1 | 8/8 | 0.321 | 0.944 | 0.996 | 0.997 | 0.833 | −0.996 | −0.997 |
| 10 | 2 | 8/8 | 0.321 | 1.278 | 1.990 | 1.991 | 1.944 | −1.988 | −1.989 |

| $N$ | $N_{sk}^{target}$ | Active MRRs (IG/OG) | $a_{IG}/a_{OG}$ | $\Delta\delta/\pi$ | $N_{sk}$ | $N_{sk}$, masked | $\Delta\delta/\pi$ (negative target) | $N_{sk}$ (negative target) | $N_{sk}$, masked (negative target) |
|---|---|---|---|---|---|---|---|---|---|
| 10 | 3 | 30/30 | 0.321 | 0.056 | 2.984 | 2.994 | 0.833 | −2.983 | −2.992 |
| 10 | 4 | 30/30 | 0.477 | 1.222 | 3.950 | 3.962 | 1.278 | −3.948 | −3.956 |
| 10 | 5 | 8/44 | 0.933 | 0.000 | 4.983 | 4.994 | 1.889 | −4.983 | −4.993 |
| 10 | 6 | 8/50 | 1.077 | 1.889 | 5.960 | 5.983 | 0.833 | −5.959 | −5.982 |
| 10 | 7 | 8/50 | 1.659 | 1.444 | 6.881 | 6.899 | 1.278 | −6.879 | −6.897 |
| 12 | 1 | 8/8 | 0.321 | 0.944 | 0.996 | 0.997 | 0.833 | −0.996 | −0.997 |
| 12 | 2 | 8/8 | 0.321 | 1.278 | 1.990 | 1.991 | 1.944 | −1.988 | −1.989 |
| 12 | 3 | 30/30 | 0.321 | 0.056 | 2.984 | 2.994 | 0.833 | −2.983 | −2.992 |
| 12 | 4 | 30/30 | 0.477 | 1.222 | 3.950 | 3.962 | 1.278 | −3.948 | −3.956 |
| 12 | 5 | 8/44 | 0.933 | 0.000 | 4.983 | 4.994 | 1.889 | −4.983 | −4.993 |
| 12 | 6 | 8/70 | 1.244 | 1.889 | 5.960 | 5.982 | 0.833 | −5.959 | −5.980 |
| 12 | 7 | 8/66 | 1.077 | 1.333 | 6.963 | 6.987 | 1.500 | −6.961 | −6.984 |
| 12 | 8 | 8/72 | 1.915 | 0.111 | 7.896 | 7.932 | 0.056 | −7.895 | −7.932 |

For positive targets, the vortex is on the OG-MRRs. Negative targets use the IG-MRRs, with the same radii of the vortex subarray and the other subarray and the same power ratio between them. The active MRRs and $a_{IG}/a_{OG}$ are listed for the positive target. For the negative target, $a_{IG}/a_{OG}$ is 1.547 divided by the listed value, and the phase bias is selected again from 36 values over 2π. Δ$\delta$ is reported modulo 2π. The masked $N_{sk}$ is calculated over the pixels with at least 3% of the maximum intensity.

## S3. Fabrication Tolerance

### *S3.A Single-MRR sensitivity*

Each geometric error is tested separately with the complete-MRR 3D-FDTD model of Section 3.1. A linewidth bias changes $W_{wg}$ and $W_{bus}$ by the same amount and the coupling gap $s$ by the opposite amount. The errors in $l_g$ and $w_g$ are equal, and a separate change in $s$ represents an additional gap error. The mesh is fixed relative to the nominal geometry. The resonance wavelength and loaded quality factor $Q_L$ are obtained by a Lorentzian fit to the complex on-axis far field ($k_x$=$k_y$=0). The efficiency $\eta_{up}$ is calculated from the far-field power and calibrated to the monitor 0.4 μm above the MRR, as in Figure 3b, because the monitor near the MRR top surface also collects part of the guided near field [Figure S5a].

The linewidth bias and $H_{wg}$ shift the two MRR resonances by similar amounts, whereas $l_g$, $w_g$, and $d_g$ produce different shifts (Figure S4 and Table S2). The gap mainly changes $Q_L$, with a resonance sensitivity of at most 0.01 nm/nm. A smaller gap increases $\eta_{up}$ but reduces the circular polarization purity.

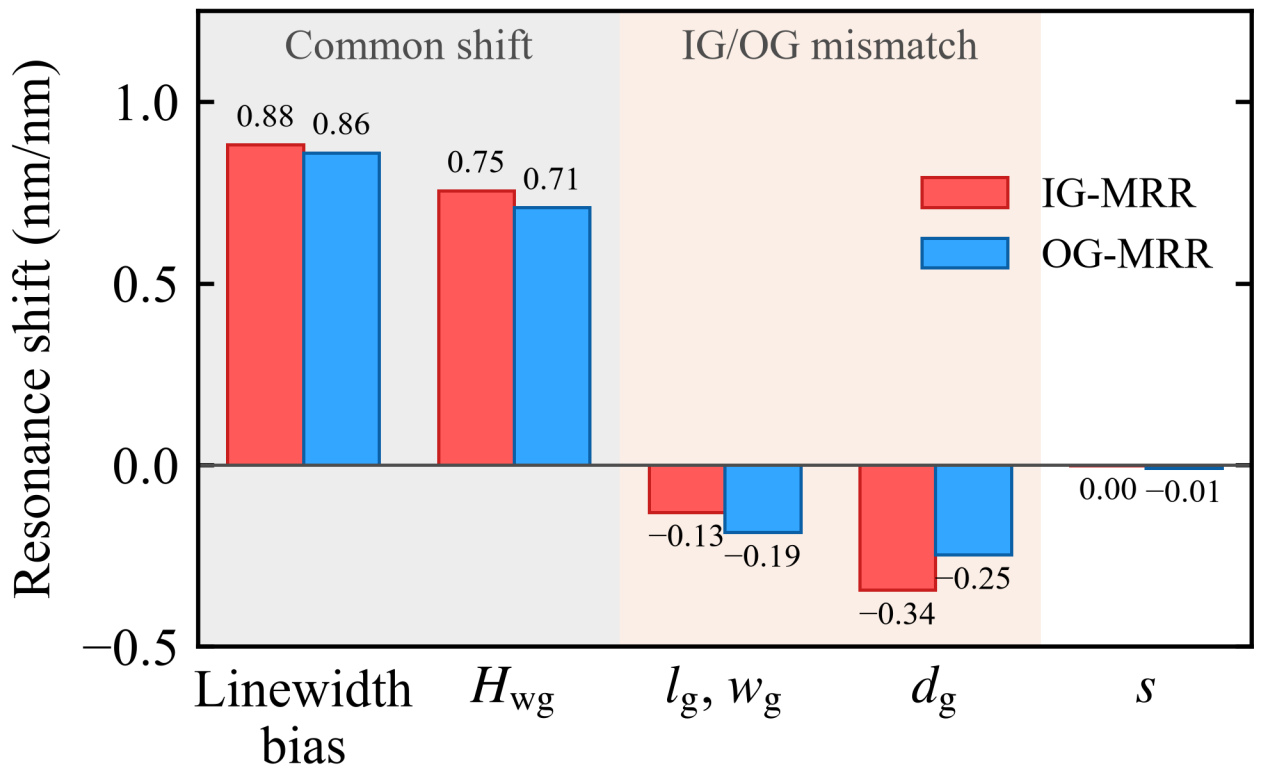

**Figure S4.** Resonance sensitivity of the IG-MRR and OG-MRR to the waveguide linewidth bias, $H_{wg}$, the common change in $l_g$ and $w_g$, $d_g$, and the additional gap change $s$. Each value is the resonance shift per nanometer of geometric change, obtained by a linear fit over the range in Table S2. "Common shift" marks errors that shift both MRR types by similar amounts, which laser tuning can follow; "IG/OG mismatch" marks errors that shift them by different amounts and separate their resonances. The complete-MRR FDTD calculations use a 30 nm mesh, except for $d_g$, which uses a 20 nm mesh.

Moving the mesh relative to the structure changes the resonance by 0.97 nm (IG) and 0.40 nm (OG) rms on the 30 nm mesh and by 0.28 nm (IG) on the 20 nm mesh. The on-axis fits give $Q_L$=875 and 711 on the 30 nm mesh, close to 840 and 708 from the efficiency spectra in Figure 3b. The Monte Carlo model uses the 20 nm mesh values, $Q_L$=912 and 767, with the OG-MRR resonance 0.60 nm above the IG-MRR resonance.

The array model uses polynomial fits of up to second order for the resonance shift, $Q_L$, and $\eta_{up}$. The contributions from separate geometric errors are added without cross terms, with lower bounds of 0.3 on $Q_L/Q_{L,0}$ and 0.1 on $\eta_{up}/\eta_{up,0}$. The fits are checked against eight combinations of geometric errors for each MRR type. The rms errors are 0.52 and 0.55 nm for the IG and OG resonances, 2.4% and 3.4% for $Q_L$, and 20% and 7.0% for $\eta_{up}$. The efficiency errors are normalized to the nominal efficiency of each MRR type. One IG combination has a resonance shift of 17.5 nm and an efficiency of 0.27 times its nominal value, compared with a fitted value of 0.82. Excluding this point reduces the IG efficiency rms error to 5.0%.

Random tooth-to-tooth changes in $l_g$ and $w_g$ are tested with standard deviations of 2 and 5 nm. The rms resonance change is at most 0.08 nm, while $Q_L$ and $\eta_{up}$ change by less than 2% and 1%, respectively. Within NA≤0.04, the target circular polarization fraction changes by less than 0.3 percentage points and the DOCP by less than 0.007. These tooth-to-tooth errors are not included in the array model.

TABLE S2

SENSITIVITY OF THE SINGLE MRRS TO THE PROCESS PARAMETERS

| Parameter | Range (nm) | Resonance shift, IG / OG (nm/nm) | $Q_L$, IG / OG | Change of $\eta_{up}$, IG / OG (%) | Change of target circular polarization fraction within NA≤0.04, IG / OG (pp) |
|---|---|---|---|---|---|
| Linewidth bias | ±10, ±20 | +0.88 / +0.86 | 875–971 / 711–891 | −32 to +8 / −15 to +6 | ≤0.9 / ≤0.6 |
| $H_{wg}$ | ±5, ±10 | +0.75 / +0.71 | 780–953 / 699–771 | −2 to +4 / −10 to +6 | ≤0.4 / ≤0.3 |
| $l_g$ and $w_g$ | ±10, ±20 | −0.13 / −0.19 | 810–976 / 640–800 | −11 to +10 / −13 to +2 | ≤0.3 / ≤0.2 |
| $d_g$ (20 nm mesh) | ±10 | −0.34 / −0.25 | 743–1106 / 706–826 | −10 to +3 / −15 to +5 | ≤0.3 / ≤0.1 |
| $s$ alone | ±10, ±20 | 0.00 / −0.01 | 741–997 / 543–882 | −17 to +17 / −11 to +2 | ≤0.8 / ≤0.2 |
| Tooth-to-tooth, rms | 2, 5 | ≤0.08 nm (rms) | ≤2% | ≤1 | ≤0.3 |

The mesh is 30 nm unless stated otherwise. Resonance sensitivities are linear fits over the listed ranges. For $H_{wg}$, the IG/OG slopes are 0.66/0.64 nm/nm over ±5 nm and 0.78/0.73 nm/nm over ±10 nm. The $Q_L$ ranges include the nominal values: 875/711 on the 30 nm mesh and 912/767 on the 20 nm mesh. All efficiency changes use far-field power from the 30 nm mesh, including the ±5 and ±10 nm tests of $d_g$. The changes are relative to the nominal efficiencies of 13.4% and 25.2%. Mesh changes affect $\eta_{up}$ by up to 8%.

### *S3.B Array model*

Each MRR contributes its nominal far field multiplied by a complex response factor. Near resonance, temporal coupled-mode theory gives the following normalized Lorentzian response in terms of wavelength [S1]:

$$L(\lambda) = \frac{1}{1 + 2iQ_L(\lambda - \lambda_{res}) / \lambda_{res}} \tag{S1}$$

where $\lambda$ is the laser wavelength, and $\lambda_{res}$ and $Q_L$ are the resonance wavelength and loaded quality factor of the MRR with geometric errors. Replacing $\lambda_{res}$ in the denominator by 1550 nm changes $L$ by less than 1%. A common wavelength shift places the nominal mean IG/OG resonance at the design wavelength of 1550 nm instead of the 1548.58 nm obtained on the 20 nm mesh. The 0.60 nm difference between the nominal resonances is retained. Laser tuning sets $\lambda$ to the mean IG/OG resonance calculated from the shared geometric errors, excluding the independent shift of each MRR.

With laser tuning, the complex weight of MRR $j$ is $w_j=w_{0,j}H_jr_j$. Here $w_{0,j}$ is the nominal input weight, $H_j=L_j(\eta_{up,j}/\eta_{up,0})^{1/2}$ is the MRR response, and $r_j$ represents residual errors in the input amplitude and phase. Let $A_t$ be the mean $|H_j|$ of subarray $t$, where $t$ denotes IG or OG. Phase calibration and amplitude-ratio adjustment replace $H_j$ by $|H_j|\min(A_{IG}, A_{OG})/A_t$. This requires the response phase of each MRR and the mean response amplitude of each subarray. Amplitude and phase calibration replaces $H_j$ by $\min(|H_j|, A_{cap})$ and also requires the response amplitude of every MRR. For each target state, the ceiling $A_{cap}$ is the largest of the 16 values $|H_j|$ for which $\Delta N_{sk}\leq 0.20$ without $r_j$. The model includes the residual errors in the input amplitude and phase but not the thermal crosstalk between the phase shifters.

The emitted power of each target state is $\Sigma_j\eta_{up,0,j}|w_j|^2$. This power is divided by its nominal value and averaged over the trials, and the reported power is the mean of the four state values. $N_{sk}$ and the type components are calculated within NA≤0.04 on a 300×300 grid. The derivatives in Eq. (12) are evaluated by finite differences, while Section S4 uses a lattice method for received fields after intensity masking. At three operating points in Table S5, trials with $\Delta N_{sk}$ within 0.05 of the threshold 0.35 are recalculated on a 600×600 grid. $N_{sk}$ changes by less than 0.01, and the lowest pass rate changes by at most 0.03 percentage points.

Equation (S1) is checked against FDTD far fields at detunings up to ±2 nm [Figure S5b–d]. Within NA≤0.04, the largest shape residual is 0.81% on the 30 nm mesh and 0.20% on the 20 nm mesh. After removal of the 1 m propagation phase and a common linear phase of approximately −4.4°/nm, the projection amplitude agrees with Eq. (S1) within 5.6% and the phase within 1.5°. In the laser-tuning calculations for Table S5, up to 29% of the MRR responses have detunings above 2 nm, and fewer than 1% have detunings above 5 nm. Beyond 5 nm, $|L|^2<0.04$ at the nominal $Q_L$. At detunings of 2.5–5 nm, 14 geometric-error cases from Table S2 give shape residuals of at most 1.8% and projection amplitudes 0.97–1.12 times the Lorentzian prediction. Within ±5 nm, the nominal on-axis spectra give amplitude ratios of 0.86–1.17 and phase differences of at most 11° relative to Eq. (S1), after removal of a linear phase.

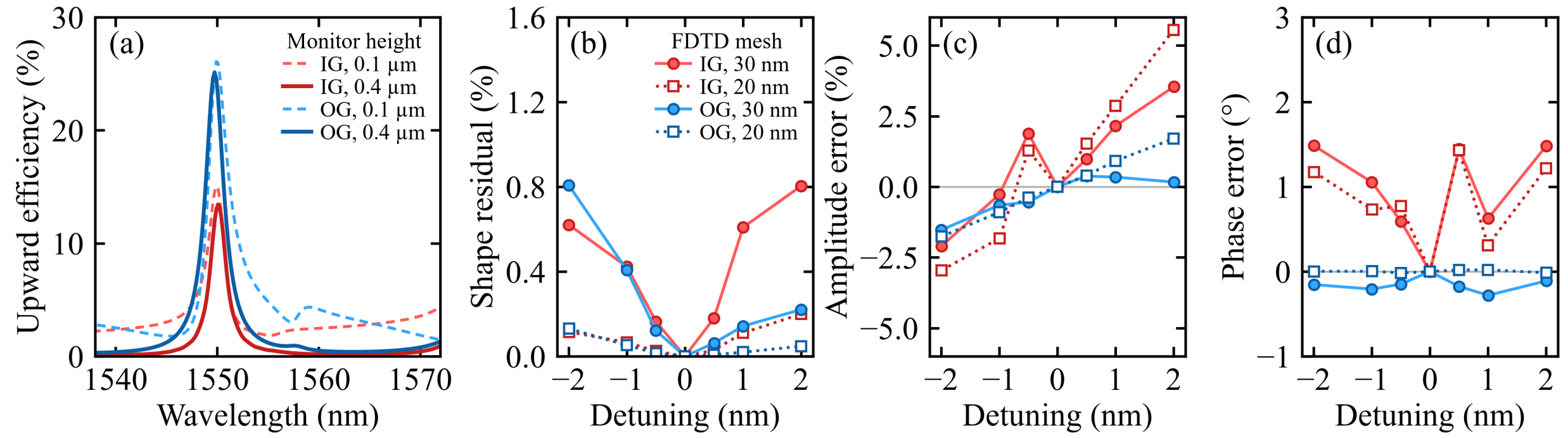

**Figure S5.** Checks of the single-MRR response used in the array model. (a) Upward emission-efficiency spectra of the nominal IG-MRR and OG-MRR. Monitors are 0.1 µm (dashed) and 0.4 µm (solid) above the MRR; the latter is used in Figure 3b. (b)–(d) Comparison of the FDTD field $\mathbf{E}$ at each detuning with the nominal field $\mathbf{E}_0$ multiplied by the Lorentzian response $L$ in Eq. (S1), within NA≤0.04. (b) Shape residual $1-|\langle \mathbf{E}_0, \mathbf{E}\rangle|^2/(\|\mathbf{E}_0\|^2\|\mathbf{E}\|^2)$. (c) Amplitude error $|c|/|L|-1$ and (d) phase error $\arg c - \arg L$ of the projection coefficient $c=\langle \mathbf{E}_0, \mathbf{E}\rangle/\|\mathbf{E}_0\|^2$. The propagation phase of the 1 m far-field projection and a common linear phase of approximately −4.4°/nm are removed. Circles and squares denote the 30 and 20 nm meshes.

The effect of geometric errors on the far-field distribution is tested by replacing the nominal fields of all IG-MRRs, all OG-MRRs, or both subarrays. The replacement fields come from the separate and combined error cases. Each field is divided by its complex projection coefficient $c$ relative to the nominal field (Figure S5), removing the overall amplitude and phase difference while retaining the spatial change. For the tested cases, $N_{sk}$ changes by at most 0.013 and the type components remain at or above 0.997. Shifting the FDTD mesh gives corresponding values of 0.014 and 0.998.

We also use the projection coefficient $c$ to evaluate the response factor across geometries. The nominal and perturbed fields are compared at the same wavelength, so the propagation phase cancels. For single-parameter cases with resonances within 5 nm of this wavelength, $c$ gives amplitude ratios of 0.94–1.10 and phase differences of at most 14° relative to $(L/L_0)(\eta_{up}/\eta_{up,0})^{1/2}$, where $L_0$ is the nominal Lorentzian response. The shape residuals are below 1.6%; mesh shifts alone give phase differences of up to 8°. After removal of the propagation and remaining common phase, the phase at resonance is fitted against the geometric errors. The IG/OG slopes are +7.9/+8.1°/nm for the linewidth bias and +3.0/+2.8°/nm for $H_{wg}$. With the shared errors in Table S3, this linear model predicts a relative-phase change of about 1° rms. The combined-error residuals are 9° rms for IG and 30° rms for OG, including one OG residual of 83°. The phase model is used only in the following sensitivity tests.

The sensitivity tests evaluate laser tuning at $\sigma_{in}$=0.25 and 0.5 nm and phase calibration at $\sigma_{in}$=0.5 and 1.5 nm for the three depth-error conditions in Table S5. They include the rms fit errors in resonance wavelength, $Q_L$, and $\eta_{up}$ from Section S3.A, projection-amplitude errors of 5% for IG and 3% for OG, and the non-Lorentzian response from the nominal on-axis spectra. Each additional error is drawn independently for the two MRR types and shared by all MRRs of the same type. The lowest pass rates change by at most 2.4 percentage points at the tested points. Adding the linear phase model with residuals of 9° for IG and 30° for OG changes these rates by at most a further 0.5 percentage points, with no further change in the phase-calibrated pass rates. The separate 0.55 nm resonance-error test in Table S5 changes the lowest pass rates by at most 4 percentage points.

Only the etch-depth samples exceed the scanned ranges, with 4.5% of the samples outside ±10 nm for a standard deviation of 5 nm. Clipping these errors to ±10 nm changes the lowest pass rate by at most 0.15 percentage points at three operating points. The lower bounds on $Q_L/Q_{L,0}$ and $\eta_{up}/\eta_{up,0}$ are not reached in the original samples.

The tolerance results apply to the geometric and resonance errors included in the array model. The model does not include sidewall angle, sidewall roughness, coupling between neighboring MRRs, or routing waveguides. Overlay errors are tested separately from the combined random errors.

### *S3.C Error levels*

Table S3 lists the errors used in Section 3.5 and their sources. The selected etch-depth errors use a reported value for silicon nitride. The Monte Carlo model draws zero-mean Gaussian errors. Geometric errors other than the etch depth are shared by all MRRs. The errors in $l_g$ and $w_g$ are equal, and the linewidth bias changes $s$ by the opposite amount. The two MRR types have different etch depths and require two shallow etches. The model assumes independent depth errors for the two etches. The reference case labeled "shared" applies the same depth error to both types while retaining their different nominal depths. Each MRR also receives an independent resonance shift with standard deviation $\sigma_{in}$ to represent local differences not described by the shared geometric errors. Overlay errors are applied as fixed displacements (Section S3.G).

The lower value of $\sigma_{in}$ is based on electron-beam-fabricated MRRs in Ref. [46]. Let $\Delta_{med}$ be the median absolute difference between the resonance wavelengths of two nominally identical MRRs. If the two resonance shifts are independent Gaussian variables with the same standard deviation, their difference has twice the variance. The relation between the median absolute deviation and the standard deviation of a Gaussian variable [S2] then gives

$$\sigma_{in} = \frac{\Delta_{med}}{\sqrt{2}\Phi^{-1}(3/4)} \approx \frac{\Delta_{med}}{0.954} \tag{S2}$$

where $\Phi^{-1}$ is the inverse cumulative distribution function of the standard normal distribution. The estimated upper value of $\sigma_{in}$ is about 1.2 nm, based on the measured linewidth and thickness variations within a chip [45] and the sensitivities in Table S2. The tests also include $\sigma_{in}$=1.5 nm to cover a larger resonance spread.

### *S3.D Errors shared by the MRRs of a chip*

Table S4 evaluates the shared geometric errors separately and in combination, with $\sigma_{in}$=0 and the same 2000 samples in each row. With the laser fixed at the nominal mean resonance, the common resonance shift reduces the relative emitted power to 0.20–0.21 for the linewidth and height errors and for the combined errors. Laser tuning keeps all four states within the skyrmion-number criterion in every trial. The remaining IG/OG mismatch changes the phase bias required for a target type. Adjusting $\Delta\delta$ restores the lowest mean type component to 0.999 but does not restore the power lost through the mismatch; further adjustment of $a_{IG}/a_{OG}$ gives little improvement in the type component. The efficiency-only case uses standard deviations of 7.6% for IG and 6.6% for OG under the shared depth-error setting. This case and the separate gap-error cases retain a 100% observed pass rate, with changes in the mean type component of at most 0.004.

Additional texture checks cover the independent 5 nm depth-error case and the two combined-error cases in Table S4. After $\Delta\delta$ is adjusted, the spatially averaged ($\cos\chi$, $\sin\chi$) vector has a mean length of at least 0.91, compared with 0.92–0.93 for the nominal states. The Stokes overlap $\Sigma S_{0,ref}(n\cdot n_{ref})/\Sigma S_{0,ref}$ with the nominal state of the same type has a mean of at least 0.994 and a fifth percentile of at least 0.972. The overlap between the nominal Néel-type and Bloch-type states is 0.29–0.37. With laser tuning alone, the mean vector length remains at 0.89–0.90 while the type component decreases. These results indicate that the mismatch mainly rotates the in-plane Stokes vectors in the tested cases.

TABLE S3

PROCESS ERRORS USED IN SECTION 3.5

| Error | Reported or estimated level | Used in Section 3.5 | Source |
|---|---|---|---|
| Linewidth bias | 3.9–5.5 nm | 5 nm | [45], [48], [49] |
| $H_{wg}$ | 0.8–2.5 nm | 2.5 nm | [45], [48], [50] |
| $l_g$, $w_g$ | not reported | 5 nm (assumed, as linewidth) | — |
| $d_g$ (per etch) | 3.8 nm (silicon nitride) | 3–5 nm | [51] |
| $s$ alone | not reported | 5 nm (assumed) | — |
| Overlay between layers | <20 nm (electron-beam lithography) | 10–30 nm (fixed displacement) | [47] |
| Intra-die resonance $\sigma_{in}$ | 0.26–1.2 nm (estimated) | 0.25–1.5 nm | [45], [46], Eq. (S2) |

The selected random-error values are standard deviations. Overlay values are fixed displacements along $\pm x$ or $\pm y$. The gap error combines the change caused by the linewidth bias with an independent additional error. Both have a standard deviation of 5 nm, giving a total gap-error standard deviation of about 7.1 nm.

### *S3.E IG/OG resonance mismatch*

The calculation in Figure S6 applies a uniform resonance mismatch between the IG and OG subarrays, with the laser at their mean resonance. All MRRs in a subarray change in the same way, so a mismatch within ±3 nm changes $N_{sk}$ by at most 0.05, while the relative phase between the subarrays changes the skyrmion type. Adjusting $\Delta\delta$ restores the target type component to approximately 1.000 at ±0.6, ±1, ±2, and ±3 nm [Figure S6b]. Two $\Delta\delta$ values separated by approximately $\pi$ can give the same type component with reversed in-plane Stokes vectors. A separate scan selects $\Delta\delta$ by maximizing the overlap with the nominal Stokes field. This gives an overlap of at least 0.995 and an averaged-vector length of at least 0.91. The results in Figure S6b use the maximum type component. The relative emitted power falls to 0.77–0.78 at 1 nm and to 0.46–0.47 at 2 nm.

TABLE S4

EFFECT OF EACH SHARED ERROR ON THE 4×4 SIGMA

| Error (standard deviation) | IG/OG mismatch, std (nm) | Fixed laser: power | Laser tuning: pass / type (5th pct.) / power | $\Delta\delta$ adjusted: type / power | $\Delta\delta$ and $a_{IG}/a_{OG}$ adjusted: type / power |
|---|---|---|---|---|---|
| Linewidth bias 5 nm, $H_{wg}$ 2.5 nm | 0.16 | 0.21 | 100% / 0.984 (0.939) / 0.99 | 0.999 / 0.99 | 0.999 / 0.96 |
| $l_g$, $w_g$ 5 nm | 0.28 | 0.70 | 100% / 0.955 (0.825) / 0.98 | 0.999 / 0.98 | 0.999 / 0.96 |
| $d_g$ 5 nm, shared | 0.50 | 0.52 | 100% / 0.877 (0.549) / 0.94 | 0.999 / 0.94 | 0.999 / 0.91 |
| $d_g$ 3 nm, independent | 1.29 | 0.68 | 100% / 0.593 (0.051) / 0.76 | 0.999 / 0.76 | 0.999 / 0.73 |
| $d_g$ 5 nm, independent | 2.15 | 0.52 | 100% / 0.522 (0.051) / 0.61 | 0.999 / 0.61 | 0.999 / 0.58 |
| $\eta_{up}$ only, changed by all errors | 0 | 0.99 | 100% / 0.999 (0.996) / 0.99 | 0.999 / 0.99 | 0.999 / 0.94 |
| $s$ alone 5 nm | 0.03 | 1.00 | 100% / 0.998 (0.993) / 1.00 | 0.999 / 1.00 | 0.999 / 0.97 |
| $s$ alone 10 nm | 0.07 | 0.99 | 100% / 0.996 (0.986) / 0.99 | 0.999 / 0.99 | 0.999 / 0.95 |
| Nominal offset only (0.60 nm) | 0 | 0.91 | 100% / 0.794 (0.750) / 0.91 | 0.999 / 0.91 | 0.999 / 0.89 |
| All errors, shared $d_g$ error | 0.59 | 0.20 | 100% / 0.694 (0.147) / 0.85 | 0.999 / 0.85 | 0.999 / 0.80 |
| All errors, independent ($d_g$ 5 nm) | 2.16 | 0.21 | 100% / 0.510 (0.045) / 0.59 | 0.999 / 0.59 | 0.999 / 0.55 |

Pass is the lowest pass rate among the four target states in Figure 5. Type is the lowest mean target component among the four Néel-type and Bloch-type states with $N_{sk}$=±1; parentheses give the lowest of their fifth percentiles. Power is the relative emitted power defined in Section S3.B, averaged over trials and then over the four states. Each row includes the efficiency changes caused by its listed errors. The efficiency-only row uses nominal resonance wavelengths and $Q_L$. Only the last three rows include the nominal 0.60 nm IG/OG resonance offset. All rows include residual errors with standard deviations of 3° in phase and 3% in relative field amplitude. For the combined errors, laser tuning ranges of ±12.8 nm (shared depth error) and ±12.4 nm (independent depth errors) around the nominal mean resonance cover 99% of the sampled mean resonance shifts.

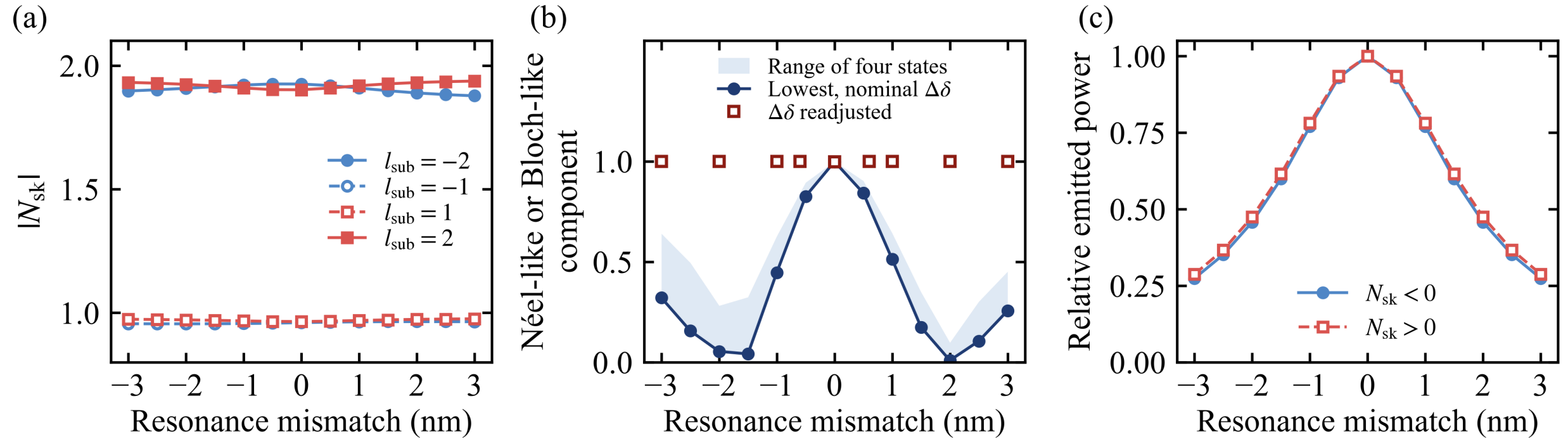

**Figure S6.** Uniform IG/OG resonance mismatch in the 4×4 SIGMA. The mismatch is $\lambda_{res,IG}-\lambda_{res,OG}$. The laser is at the mean resonance, with $Q_L$=912 for the IG-MRR and 767 for the OG-MRR. (a) $|N_{sk}|$ of the four states in Figure 5. Blue circles and red squares denote negative and positive targets; solid and dashed lines denote $|l_{sub}|$=2 and 1. (b) Target type components of the four states in Figure 4 and Figure S1. Circles give the lowest component at the nominal $\Delta\delta$, and shading gives the range across the four states. Squares give the lowest component after phase adjustment at the tested mismatches, with the nominal result included at zero mismatch. (c) Relative emitted power for negative and positive targets.

### *S3.F Independent resonance shifts within a chip*

Figure S7 and Table S5 combine the shared geometric errors with an independent resonance shift for each MRR. The standard deviation of this shift is $\sigma_{in}$, and each sampled condition uses 4000 trials. Failures occur mainly in the states with $|l_{sub}|$=2. After the input phase of each MRR is calibrated and $a_{IG}/a_{OG}$ is adjusted, the states with $|l_{sub}|$=1 pass in at least 96% of trials up to $\sigma_{in}$=1.5 nm.

Amplitude and phase calibration uses the known complex response of each MRR to correct the response phases and to reduce the strongest response amplitudes to a ceiling (Section S3.B). The ceiling keeps the calculated $\Delta N_{sk}$ of each target state within 0.20, which leaves a margin to the criterion of 0.35 for the assumed residual errors in the input amplitude and phase, with standard deviations of 3% and 3°. The mean number of attenuated MRRs is below 0.1 at $\sigma_{in}$=0.25 nm and 3.0–3.3 at 1.5 nm. The lowest pass rate is at least 99.9% in all cases of Table S5. The emitted power decreases as $\sigma_{in}$ increases. It can exceed the power obtained with phase calibration, which also attenuates one subarray to adjust $a_{IG}/a_{OG}$. For both calibration methods, the lowest mean target type component is at least 0.997 and the lowest fifth percentile is at least 0.989 across the conditions in Table S5, compared with a lowest mean of 0.54–0.82 for laser tuning alone.

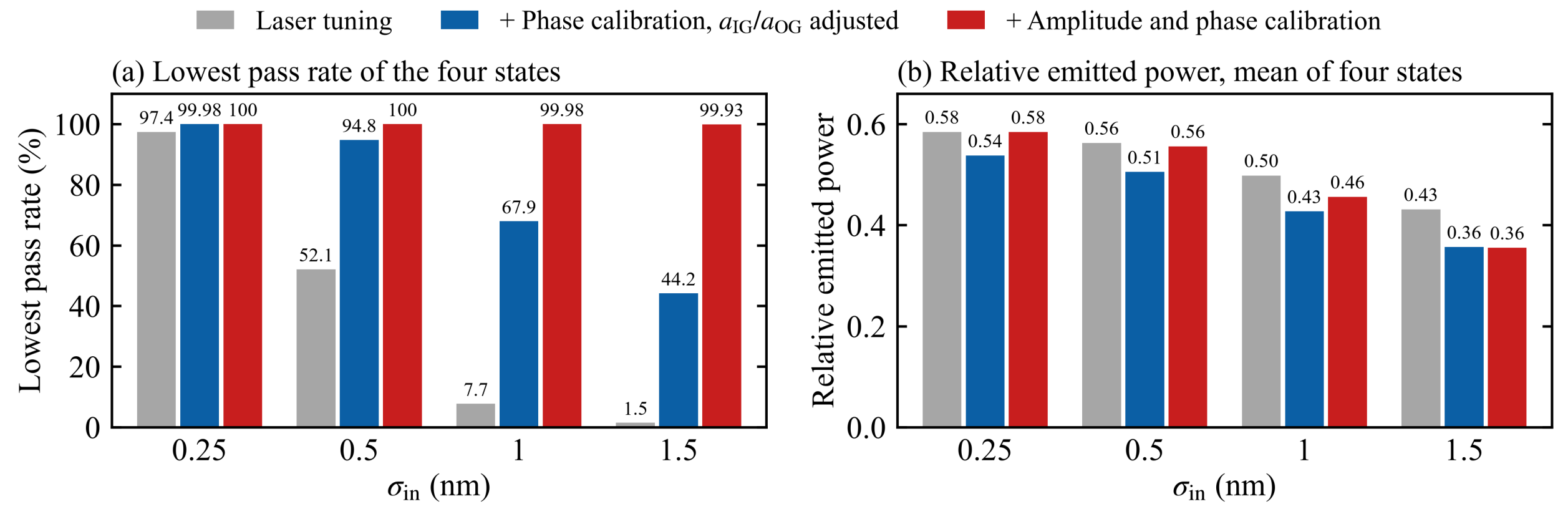

**Figure S7.** Monte Carlo results for shared geometric errors and independent MRR resonance shifts in the 4×4 SIGMA. Each sampled condition uses 4000 trials. The nominal $Q_L$ values are 912 and 767. Residual phase and relative field-amplitude errors have standard deviations of 3° and 3%. The depth errors of the two shallow etches are independent with a standard deviation of 5 nm; Table S5 lists the other depth conditions. (a) Lowest pass rate among the four states in Figure 5. (b) Relative emitted power averaged over trials and then over the four states. Gray bars use laser tuning; blue bars also include calibration of the input phase of each MRR and adjustment of $a_{IG}/a_{OG}$; red bars include calibration of the input amplitude and phase of each MRR. The Wilson 95% intervals of the pass rates have half-widths of at most 1.6 percentage points.

TABLE S5

LOWEST PASS RATE OF THE FOUR STATES AND MEAN RELATIVE EMITTED POWER (IN PARENTHESES) VERSUS $\sigma_{in}$

| Calibration | Depth error ($d_g$ std) | $\sigma_{in}$=0.25 nm | 0.5 nm | 1 nm | 1.5 nm |
|---|---|---|---|---|---|
| Laser tuning | Shared (5 nm) | 94.1% (0.82) | 38.4% (0.75) | 3.6% (0.60) | 0.5% (0.49) |
| | Independent (3 nm) | 96.8% (0.70) | 45.4% (0.66) | 4.4% (0.56) | 0.8% (0.47) |
| | Independent (5 nm) | 97.4% (0.58) | 52.1% (0.56) | 7.7% (0.50) | 1.5% (0.43) |
| Phase calibration | Shared (5 nm) | 100% (0.77) | 99.0% (0.69) | 83.5% (0.53) | 55.2% (0.41) |
| | Independent (3 nm) | 100% (0.66) | 96.6% (0.60) | 75.7% (0.49) | 50.4% (0.39) |
| | Independent (5 nm) | 99.98% (0.54) | 94.8% (0.51) | 67.9% (0.43) | 44.2% (0.36) |
| Phase calibration, model error 0.55 nm | Shared (5 nm) | — | 98.0% (0.64) | 79.5% (0.51) | — |
| | Independent (5 nm) | — | 95.0% (0.49) | 67.4% (0.42) | — |
| Amplitude and phase calibration | Shared (5 nm) | 100% (0.82) | 100% (0.75) | 100% (0.57) | 99.95% (0.42) |
| | Independent (3 nm) | 100% (0.70) | 100% (0.66) | 99.98% (0.52) | 99.95% (0.39) |
| | Independent (5 nm) | 100% (0.58) | 100% (0.56) | 99.98% (0.46) | 99.93% (0.36) |

Each condition uses 4000 trials; phase calibration includes adjustment of $a_{IG}/a_{OG}$. The model-error tests add a resonance offset with a standard deviation of 0.55 nm, drawn for each MRR type and shared within that type. Power is normalized to the nominal design and averaged over trials and then over the four states. The Wilson 95% intervals of the pass rates have half-widths of at most 1.6 percentage points (94.0–95.4% for 94.8%). A pass rate of 100% means no failure in 4000 trials, with a lower confidence bound of 99.90%. With amplitude and phase calibration, the lowest fifth percentiles of the four states' relative emitted power are 0.62, 0.55, 0.24, and 0.12 for the shared depth error and 0.14, 0.15, 0.12, and 0.08 for independent depth errors of 5 nm, in the $\sigma_{in}$ order of the table.

### *S3.G Overlay error*

An overlay error shifts the gratings relative to the ring waveguide. The resonance shift is less than 0.15 nm, and the main effect is a change in the IG-MRR far field [Figure S8a]. The array calculation replaces the fields of all IG-MRRs or all OG-MRRs with fields calculated for a grating displacement along $\pm x$ or $\pm y$. The change in $N_{sk}$ is at most 0.05 [Figure S8c]. The lowest type component falls to 0.83 at 10 and 20 nm and to 0.59 at 30 nm, but returns to approximately 0.999 after $\Delta\delta$ and $a_{IG}/a_{OG}$ are adjusted [Figure S8b]. After the adjustment, in which no MRR is driven above its nominal input, the relative emitted power of the array is at least 0.96, 0.92, and 0.80 at 10, 20, and 30 nm, and all four states in Figure 5 remain within the skyrmion-number criterion.

The Stokes-field overlap and averaged-vector length quantify the remaining changes in the local texture. After $a_{IG}/a_{OG}$ is adjusted, a separate scan selects $\Delta\delta$ by maximizing the overlap with the nominal Stokes field. The overlap is at least 0.990 for a 20 nm displacement and 0.956 for a 30 nm displacement. For IG-MRR displacements, the averaged-vector length decreases from the nominal range of 0.84–0.94 to 0.80 and 0.73, respectively. The OG-MRR results use the 30 nm mesh and do not change monotonically with displacement. Overlay reduces the target circular polarization fraction of the IG-MRR, whereas $N_{sk}$ of the tested arrays changes by at most 0.05.

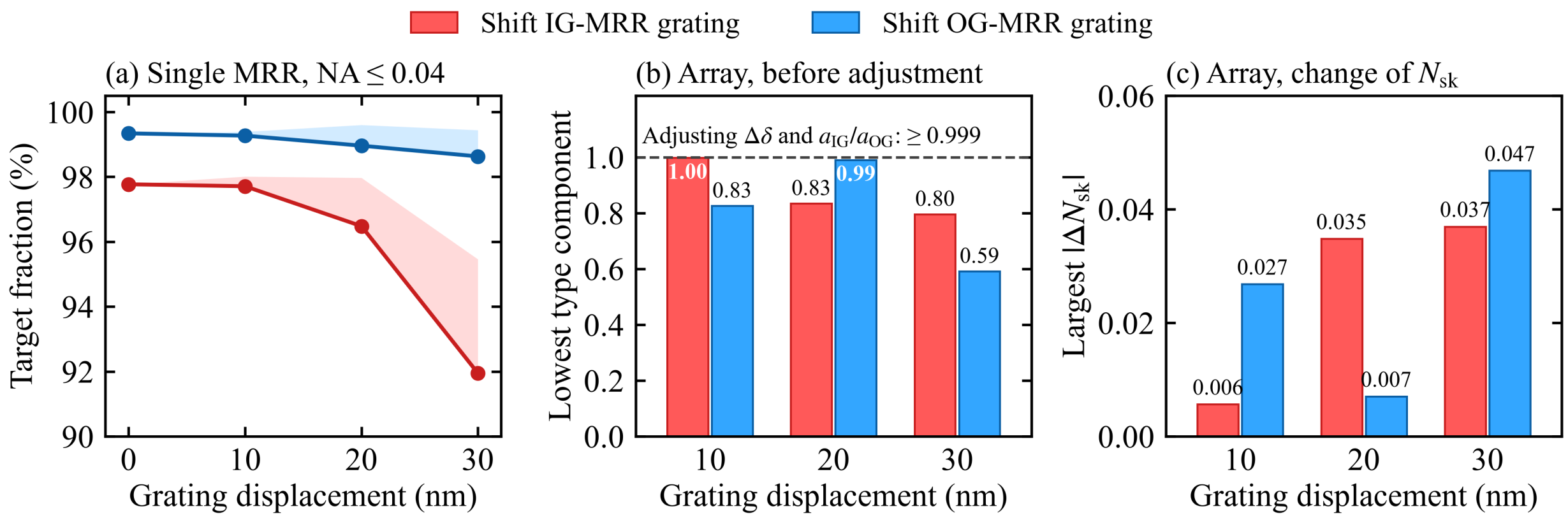

**Figure S8.** Effect of a displacement between the grating and waveguide layers. Each displacement is applied along $+x$, $-x$, $+y$, and $-y$, and each quantity is shown for its worst direction. (a) Target circular polarization fraction within NA≤0.04; the shading shows the range over the four directions. The IG-MRR uses the 20 nm mesh and the OG-MRR the 30 nm mesh. (b) Lowest target type component among the four type-switched states of the 4×4 SIGMA at the nominal $\Delta\delta$ and $a_{IG}/a_{OG}$. The displaced fields replace all IG-MRR fields (red) or all OG-MRR fields (blue). The dashed line marks the lowest value after $\Delta\delta$ and $a_{IG}/a_{OG}$ are adjusted. (c) Largest change in $N_{sk}$ among the four states in Figure 5 at the nominal input settings.

### *S3.H Modulation errors and element failures*

Table S6 evaluates modulation errors and MRR failures without resonance errors, where a failed MRR is assumed to emit no light. The states with $|l_{sub}|=1$ pass in at least 99.97% of the trials for all tested modulation errors, whereas the pass rates of the states with $|l_{sub}|=2$ decrease when the phase error increases from 10° to 15°. Every single-MRR failure leaves all four states within the skyrmion-number criterion. When two MRRs fail, the states with $|l_{sub}|=1$ still pass for all pairs, and only the states with $|l_{sub}|=2$ fail for some of the pairs.

TABLE S6

MODULATION ERRORS AND ELEMENT FAILURES WITHOUT RESONANCE ERRORS

| Condition | Trials | Pass rate, $l_{sub} = -2 / -1 / 1 / 2$ (%) |
|---|---|---|
| Phase 10°, amplitude 10% | $10^4$ | 99.99 / 100 / 100 / 99.57 |
| Phase 15°, amplitude 10% | $10^4$ | 95.64 / 99.97 / 99.98 / 90.00 |
| Amplitude 20% only | 2000 | 99.65 / 100 / 100 / 96.9 |
| One failed MRR | all 16 | 100 / 100 / 100 / 100 |
| Two failed MRRs | all 120 | 66.7 / 100 / 100 / 63.3 |

The listed phase and relative field-amplitude errors are standard deviations. For errors of 10° and 10%, each of the four type-switched states has a mean type component of at least 0.990 and a fifth percentile of at least 0.961.

## S4. TRANSMITTER AND RECEIVER OF THE SKYRMION CHANNEL

The SIGMA far field within NA≤0.04 is mapped to the transmitter plane using $x=f\cdot k_x/k_0$ and $y=f\cdot k_y/k_0$. A circular stop of radius $0.04f$ limits the transmitted field. The focal length $f$=1.232 m gives a mean rms radius of approximately 3 cm for the four SIGMA beams. The stop diameter is 9.86 cm, smaller than the 15 cm transmitting aperture used for both channels in Section 4. The rms radius is calculated from the second moment of the intensity. Each LG beam has an rms radius of 3 cm and the same power as a SIGMA beam. The 100 m propagation uses a 2048×2048 grid over a 1.2 m×1.2 m window, with a sampling interval of 0.586 mm. Table S7 gives the beam radii and the receiver reference values without turbulence.

The receiver aperture has a diameter of 15 cm. The intensity maximum of each received field is calculated within this aperture. The mask keeps pixels with at least 3% of this maximum and removes low-intensity regions where the normalized Stokes vector $\mathbf{n}$ is sensitive to numerical noise. Equation (12) is evaluated on the retained pixels using the Berg–Lüscher lattice method [S3], [S4]. Each square cell with four retained corners is divided into two triangles. The skyrmion number is

$$N_{sk} = \frac{1}{4\pi}\sum_{cells}\left[\Omega(n_1,n_2,n_3)+\Omega(n_1,n_3,n_4)\right]$$
$$\Omega(\mathbf{a},\mathbf{b},\mathbf{c}) = 2\mathrm{atan2}[\mathbf{a}\cdot(\mathbf{b}\times\mathbf{c}),1+\mathbf{a}\cdot\mathbf{b}+\mathbf{b}\cdot\mathbf{c}+\mathbf{c}\cdot\mathbf{a}] \tag{S3}$$

where $\mathbf{n}_1$–$\mathbf{n}_4$ are the normalized Stokes vectors at the four cell corners in counterclockwise order. The function Ω gives the signed solid angle of three Stokes vectors on the Poincaré sphere. Equation (S3) adds the solid angles of the triangles without calculating spatial derivatives of $\mathbf{n}$. The mask is calculated separately for each reference field and each turbulent trial using the same relative intensity threshold.

TABLE S7

TRANSMITTED BEAMS AND RECEIVED REFERENCES WITHOUT TURBULENCE

| Symbol | SIGMA rms radius (cm) | SIGMA $N_{sk}$ at transmitter | SIGMA $N_{sk}$ reference at receiver | Captured power, 15 cm aperture | LG rms radius (cm) |
|---|---|---|---|---|---|
| −2 | 2.98 | −1.925 | −1.834 | 99.7% | 3.00 |
| −1 | 3.04 | −0.959 | −0.929 | 99.8% | 3.00 |
| +1 | 3.03 | 0.964 | 0.936 | 99.8% | 3.00 |
| +2 | 2.92 | 1.901 | 1.774 | 99.7% | 3.00 |

The transmitter stop has a radius of $0.04f$=4.93 cm at the back focal plane, with $f$=1.232 m. Receiver references are the $N_{sk}$ values after 100 m of propagation without turbulence, calculated with the intensity mask and Eq. (S3). The transmitter values are also calculated with Eq. (S3), on the propagation grid after the far field is mapped to the transmitter plane and with a relative intensity threshold of $10^{-6}$, so they can differ from the values in Figure 5 in the last digit. The receiver aperture captures more than 99.99% of the power of each LG beam.

## REFERENCES

Reference numbers without "S" refer to the reference list of the main article.